\documentclass[preprints,article,accept,pdftex,moreauthors]{Definitions/mdpi} 
\firstpage{1} 
\pubvolume{1}
\issuenum{1}
\articlenumber{0}
\pubyear{2022}
\copyrightyear{2022}
\externaleditor{Academic {Editor
}: Dmitri Donetski }
\datereceived{4 May 2022} 
\dateaccepted{3 June 2022} 
\datepublished{28 June 2022} 
\hreflink{https://doi.org/} 

\Title{How to Recognize the Universal Aspects of Mott Criticality?}

\TitleCitation{How to Recognize the Universal Aspects of Mott Criticality?~}

\Author{Yuting Tan
 $^{1,}$*, Vladimir Dobrosavljevi\'{c} $^{1,}$* and Louk Rademaker $^{2,}$*}

\AuthorNames{Yuting Tan, Vladimir Dobrosavljevi\'{c} and Louk Rademaker}

\AuthorCitation{Tan, Y.; Dobrosavljevi\'{c}, V.; Rademaker, L.}

\address{%
$^{1}$ \quad Department of Physics and National High Magnetic Field Laboratory,
Florida State University, \linebreak Tallahassee, FL 32306, USA \\
$^{2}$ \quad Department of Theoretical Physics, University of Geneva, 1211 Geneva, Switzerland}

\corres{Correspondence: ytan@magnet.fsu.edu (Y.T.); vlad@magnet.fsu.edu (V.D.); louk.rademaker@gmail.com (L.R.)}

\abstract{In this paper we critically discuss several examples of two-dimensional electronic systems displaying interaction-driven metal-insulator transitions of the Mott (or Wigner--Mott) type, including dilute two-dimension electron gases (2DEG) in semiconductors, Mott organic materials, as well as the recently discovered transition-metal dichalcogenide (TMD) moir\'e bilayers.  Remarkably similar behavior is found in all these systems, which is starting to paint a robust picture of Mott criticality.
Most notable, on the metallic side a resistivity maximum is observed whose temperature scale vanishes at the transition.
We compare the available experimental data on these systems to three existing theoretical scenarios: spinon theory, Dynamical Mean Field Theory (DMFT) and percolation theory.
We show that the DMFT and percolation pictures for Mott criticality can be distinguished by studying the origins of the resistivity maxima using an analysis of the dielectric response.}

\keyword{Mott transition; quantum criticality; resistivity maxima; dielectric response;
dilute 2DEGs; Mott organics; twisted transition-metal dichalcogenide bilayers; dynamical mean field theory; percolation theory; spinon theory} 

\begin{document}


\section{Introduction}

The physics of strongly correlated matter has many faces. Still, for a majority of systems the underlying theme is the role of ``{\em Mottness}''~\cite{Phillips:2007}. 
It is clear that if one aspect of strong correlations should be understood first, it should be the fundamental nature of the Mott metal-insulator transition~\cite{mott1990metal}. 
Its simplest reincarnation is the transition induced by tuning the bandwidth at half-filling, a setup that produced rather spectacular advances in recent years. 
Several systems were identified as nearly-ideal realizations of this paradigm, allowing systematic study using a wide arsenal of experimental probes. 

In this article we present an overview of three classes of two-dimensional experimental systems that exhibit bandwidth-controlled Mott criticality: 
dilute two-dimensional electron gases in 
semiconductors, ``Mott-organic'' compounds, and transition-metal dichalcogenide moir\'{e} systems. Thereby we aim to present the experimental facts as ``bland'' as possible, in Section~\ref{Sec:Experiments}, without favoring one or the other theoretical explanation. The remarkable similarities between these model systems suggests a robust universality, including characteristic behavior such as the appearance of resistivity maxima.

Possible explanations of two-dimensional Mott criticality follow in the section thereafter (Section~\ref{Sec:Theory}), where the experimental distinguishable features of each theory takes the forefront. This is followed by a separate discussion of the largely-overlooked utility of dielectric spectroscopy in Section~\ref{Sec:Dielectric}, in not only identifying phase segregation and spatial inhomogeneity, but also in revealing the thermal destruction of coherent quasiparticles associated with Landau's Fermi liquid theory. 

However, first we need to address the demarcation of our topic. What makes the metal-to-insulator transition in these systems stand out from `traditional' metal-to-insulator transitions~\cite{Imada:1998er, dobrosavljevic2012conductor}?


\section{In Search of Mott Criticality}
\label{Sec:IntroMott}

Condensed matter physics, or recently for sales purposes re-branded as ``quantum matter physics'', is the study of electric and magnetic properties of materials that surround us. The grand question that trumps all others is: how to understand which materials conduct electricity and which are insulating? Traditionally, a conducting material is called a {\em metal}---not to be confused with the chemical, metallurgical or astronomical meaning of that word. Our main question (metal or insulator?) has not only tremendous technological applications (in fact, all modern electronic technology depends on our ability to rapidly switch materials between metallic and insulating behavior), but also requires a thorough understanding of the problem of emergent behavior of many interacting quantum-mechanical electrons and ions.

Only at zero temperature does there exist a sharp difference between insulators and metals~\cite{dobrosavljevic2012conductor}. There are three distinct possibilities: zero conductivity $\sigma(T=0) = 0$ means insulating; zero resistivity $\rho(T=0) = 0$ means a superconductor; and anything in between is a metal $\sigma(T=0) = 1/\rho(T=0) \neq 0$. At any nonzero temperature, an insulator typically has activated behavior $\rho(T) \sim e^{\Delta/T}$ whereas the standard Fermi liquid theory of a metal predicts a temperature-squared increase of the resistivity $\rho(T) = \rho_0 + A T^2$. It has therefore become common-place to use the {\em derivative} of the resistivity $d\rho/dT$ as a measure of whether something is conducting ($d\rho/dT>0$) or insulating ($d\rho/dT<0$)---but this is highly misleading! 
As we will show later in Section~\ref{Sec:Experiments}, close to a Mott metal-insulator transition we often find non-monotonic behavior of the resistivity as a function of temperature, making the `derivative' criterion useless. 
Even worse, there exist cases where the resistivity has $d\rho/dT<0$ but at zero temperature it does not diverge, signalling that this is not a true insulator (see e.g.,~\cite{Cao:2018kn,10.1021/acs.nanolett.1c03066}). 
Another example is the case of Mooij correlations~\cite{10.1002/pssa.2210170217,Ciuchi:2018cd}, where the temperature-derivative of the resistivity in a metal can become negative.
Consequently, since only at zero temperature the insulator/metal distinction is well-defined, we must stick with that definition. Regardless of the slope, a material is a metal if its resistivity {\em does not diverge} as $T \rightarrow 0$.

Many materials can be understood within the framework of {\em band theory} and its extensions such as Fermi liquid, Boltzmann transport, and density functional theory. This framework provides a very simple answer to the metal-or-insulator question: if the Fermi level lies in the middle of a band gap, the system is insulating; otherwise, the system behaves as a metal. This concept has the important consequence that for a crystalline material with (up to some weak disorder) well-defined unit cells, insulators can only appear when there is an even number of electrons per unit cell. Consequently, within the band theory picture, there exist only three possible routes to induce a metal-to-insulator transition: by changing the electronic density; via spontaneous symmetry breaking; or via band overlap when the filling is even. An example of the first is doping a semiconductor, which is the metal-insulator transition we induce on a daily basis inside transistors. An example of the second is the transition into antiferromagnetic ordering: when the system is at half-filling of a band (meaning one electron per unit cell), after antiferromagnetic unit cell doubling there are two electrons per unit cell, and the system can become a band insulator. The third case can be realized by for example straining a system such that the band gap changes from positive to negative.

There are, however, two main exceptions to the paradigm of band theory. On the one hand, disorder can become so large as to prevent the motion of the charge carriers---this is known as Anderson localization~\cite{Evers:2008}. On the other hand, the presence of very strong electron-electron interactions can force the electrons to become ``stuck'' like in a traffic jam---this is known as {\em Mott insulation}~\cite{mott1990metal}. The standard model of Mott insulation is the Hubbard model with a tight-binding Hamiltonian:
\begin{linenomath}
\begin{equation}
    H = - t \sum_{\langle ij \rangle \sigma} c^\dagger_{i\sigma} c_{j\sigma} + U \sum_{i} n_{i \uparrow} n_{j \downarrow},
\end{equation}
\end{linenomath}
where $t$ is the nearest-neighbor hopping on some lattice and $U$ is the onsite repulsion. When $U=0$, the system is a metal when half-filled. When $U \gg t$, it becomes energetically favorable to occupy each site with exactly one electron rather than to fill bands up to the Fermi level. The resulting Mott state can therefore not be described by band theory!

Mott insulators have been observed in a wide variety of materials, most famously transition-metal oxides, including high $T_c$ superconducting cup rates~\cite{Imada:1998er}. Observing a clear transition from a standard Fermi liquid metal to a Mott insulator, however, is quite elusive. This transition can be induced either by changing the electronic density (``filling-controlled'') or by changing the ratio $U/t$ (``bandwidth-controlled''). The filling-controlled Mott transition~\cite{Lee:2006de} notoriously leads to a whole zoo of different instabilities, pseudogaps, and strange metal behavior, and is typically masked at low temperatures by superconductivity. The bandwidth-controlled Mott transition is, in contrast, often masked by (antiferromagnetic) spin order that hides any Mottness behind the veil of unit cell doubling.

This might, at first, suggest that {\em Mott criticality} is something unattainable. By ``criticality'' we mean that approaching the Mott transition we find vanishing energy scales, and that the resistivity curves display scaling behavior. There are, however, two clever tricks to realize Mott criticality. The first trick is {\em dimensionality}: a transition that is strongly first-order in $d=3$ dimensions often becomes continuous or weakly first-order in $d=2$ dimensions. The most striking example of this is, of course, the solidification of $^3$He. The second, and perhaps even more important trick is {\em frustration}: if the lattice structure is highly frustrated (with competing magnetic interactions) one can avoid~\cite{Balents:2010ds} antiferromagnetic ordering altogether---revealing the true Mott transition.

In this review we, therefore, focus on three classes of systems that are indeed (quasi) two-dimensional as well as frustrated: Wigner crystals in extremely dilute two-dimensional electron gases; layered Mott organic compounds; and the more recent addition of transition-metal dichalcogenide (TMD) moir\'{e} bilayers. Indeed, as we will show in Section~\ref{Sec:Experiments}, these systems all seem to exhibit remarkably similar distinct features, including clear signatures of critical resistivity scaling. Because these systems all have a fixed electron density per unit cell of $n=1$ (at least in the insulating limit), the observed transitions are plausibly within the universality class of bandwidth-tuned Mott transitions.

A brief side-note is in order: we briefly mentioned superconductivity and disorder-induced insulators. These phases can also have a continuous transition between them, the so-called superconductor-to-insulator transition ~\cite{Sondhi:1997,Saito:2016ht}. This, however, is an interesting topic that falls outside the scope of this review. Similarly, we also will not consider disorder-driven metal-insulator transitions~\cite{Ramakrishnan:1985,Belitz:1994}, since this regime typically does not include any Mottness. More general but also somewhat older reviews of metal-insulator criticality can be found in Refs.~\cite{Imada:1998er,Abrahams:2001fw,Spivak2010RvMP,dobrosavljevic2012conductor,kravchenko2017strongly}.

\section{Experiments} 
\label{Sec:Experiments}

Given that experimental results should always be leading, the aim of this section is to introduce three material systems that are likely exhibiting a bandwidth-tuned Mott metal-insulator transition: dilute 2DEGs, organics, and moir\'{e} systems. To support the clarity of interpretation, we will stress {\em experimental} similarities between these systems without much room for theoretical guesswork---that is the next section's realm.

\begin{figure}[H]

\includegraphics[width=5.4in]{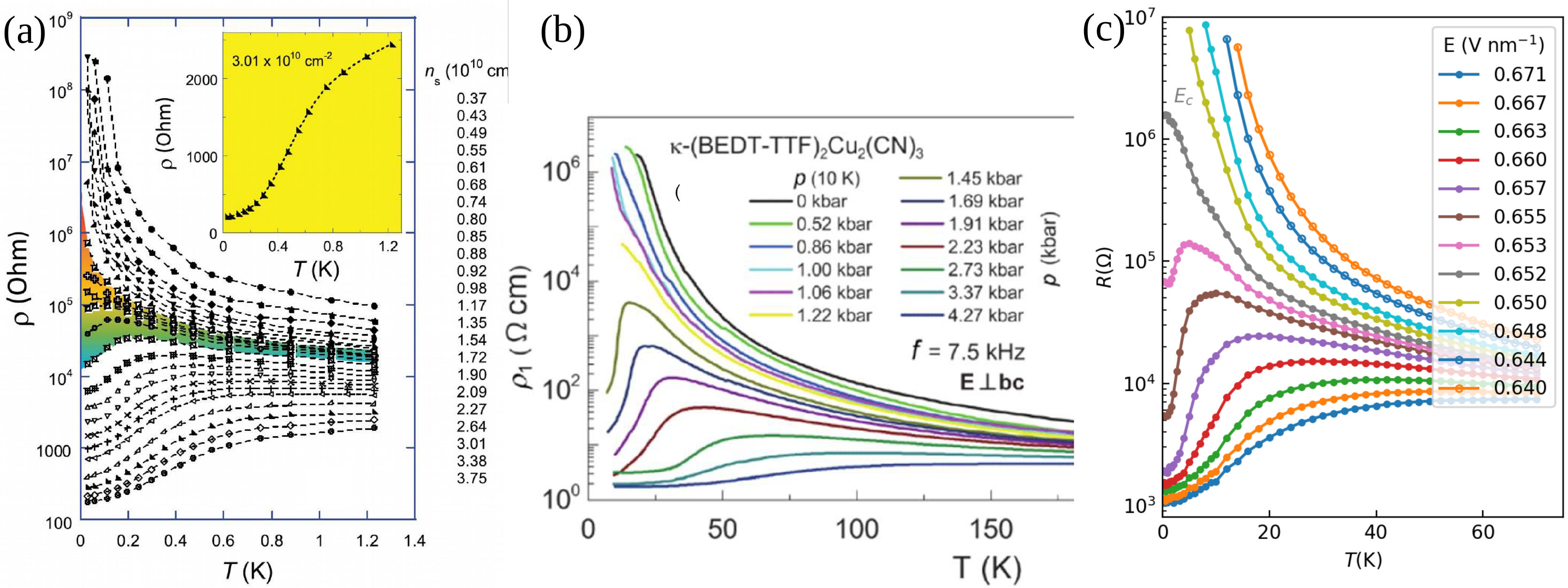}
\caption{The key observable revealing a metal-insulator transition is the resistivity. Here we show $\rho$ vs. $T$ resistivity curves as a function the tuning parameter, for representative examples of the three material systems considered: (\textbf{a}) 2DEG in Si-MOSFET tuned by electronic density~(reprinted with permission from Ref. \cite{kravchenko2019} Copyright 2019 American Physical Society), (\textbf{b}) Mott organic material $\kappa\text{-(BEDT-TTF)}_2\text{Cu}_2\text{(CN)}_3$ tuned by pressure~\cite{Pustogow:2021npj}, and (\textbf{c}) TMD moir\'{e} bilayer MoTe$_2$/WSe$_2$ tuned by displacement field (Data imported from~\cite{Li2021ContinuousMT}). In all cases, one observes distinct resistivity maxima on the metallic side, at a temperature $T_{max}$ that decreases towards the transition.
}
\label{Fig:Resistivity}
\end{figure}

\begin{figure}[H]
\includegraphics[width=0.96\textwidth]{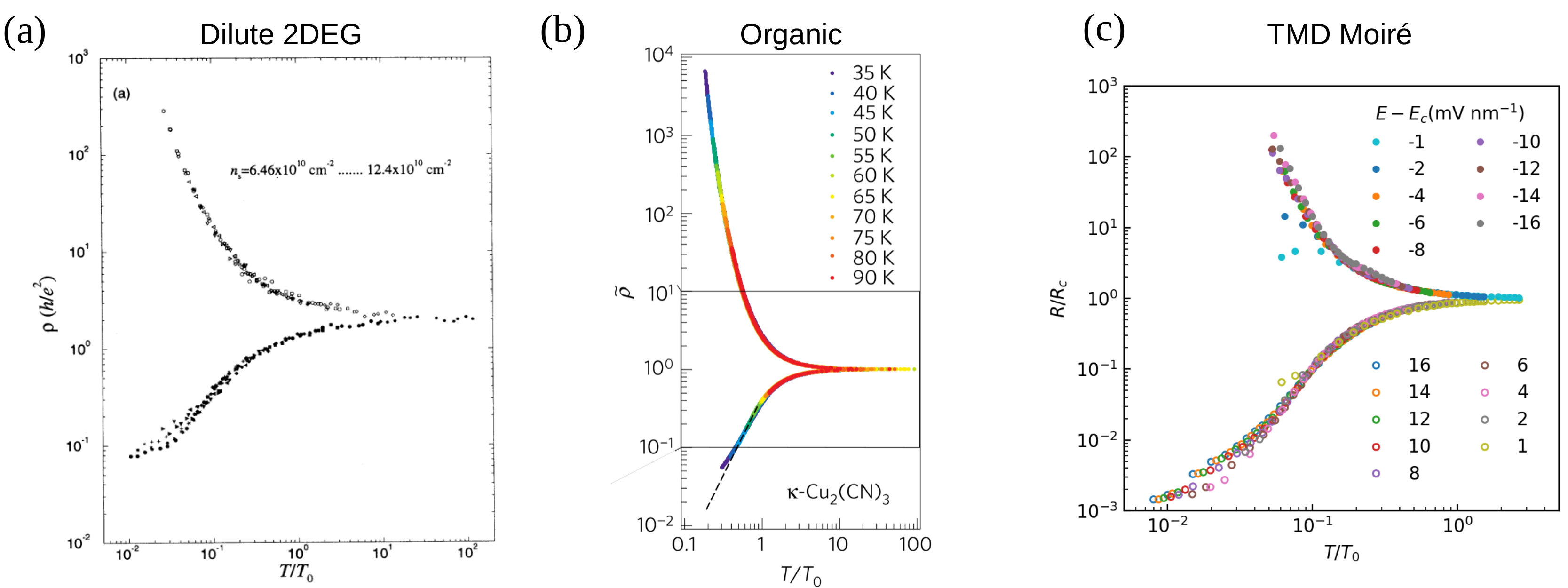}
\caption{Critical scaling has been observed in all three experimental systems, when the resistivity is plotted versus $T/T_o$ where $T_o$ is the characteristic crossover (quantum critical) energy scale. Note that in all cases a strong ``mirror'' symmetry~\cite{kravchenko1997mirror,gang41997mirror} exists between the insulating (upper) and metallic (lower) scaling branch.
(\textbf{a}) In a dilute 2DEG, scaling of the bare resistivity $\rho(T)$ was achieved by simply rescaling $T$ with $T_o \sim |\delta|^{1.6}$~(Adapted with permission from Ref. \cite{Kravchenko:1995PRB} Copyright 1995 American Physical Society); 
(\textbf{b}) In organic compounds, the normalized resistivity $\tilde{\rho}$ is obtained by normalizing the resistivity by the critical resistivity along the Widom line. This leads to excellent scaling collapse with $T_o \sim |\delta|^{0.60\pm 0.01}$~(Adapted with permission from Ref. \cite{Furukawa:2015eo} Copyright 2015 Springer Nature);
\mbox{(\textbf{c}) A} similar approach was followed in TMD moir\'{e} bilayer MoTe$_2$/WSe$_2$, with similar $T_o \sim |\delta|^{0.70\pm0.05}$ (Data imported from~\cite{Li2021ContinuousMT}). 
}
\label{Fig:QCscaling}
\end{figure}

\textls[-15]{While each system has a different tuning parameter (density, pressure, or field), the {\em electrical resistivity} through the transition is the key observable, see Figure~\ref{Fig:Resistivity}. Its behavior reveals how the transport gap $\Delta$ decreases when we approach the transition from the insulating side; as well as how the resistivity behaves on the metallic side, where Fermi liquid behavior \linebreak $\rho(T) = \rho_0 + AT^2$ is typically seen at $T < T_{FL}$ with an enhanced effective mass $m^*$. Remarkably, in all systems one also observes distinct {\em resistivity maxima} at $T \sim T_{max} > T_{FL}$, signalling the breakdown of coherent transport. Crossover to the quantum critical regime is described by an additional temperature scale $T_o$, which is extracted from the scaling collapse of the resistivity curves as shown in Figure~\ref{Fig:QCscaling}. }


A practical summary of the experimental results is presented in Table~\ref{Tab:Experiments}.

\begin{table}[H]
\caption{A summary of available experimental results for the three classes of systems considered. The sources (references) are given in the text below. Question-marks indicate the lack of reliable data. Fermi liquid ($T^2$) transport behavior has not been documented in 2DEG systems, in contrast to strong evidence for it in Mott organics and TMD moir\'e bilayers. Note that the characteristic energy scales $\Delta$, $(m^*)^{-1}$, $T_{FL}$, $T_{max}$, as well as $T_o$ display similar continuous decrease towards the transition in all three systems, consistent with general expectations for quantum criticality. One should keep in mind that the error bars on the estimated exponent could be substantial, since the results typically depend strongly on the utilized fitting range.\label{Tab:Experiments}}
\newcolumntype{C}{>{\centering\arraybackslash}X}
\begin{tabularx}{\textwidth}{CCCC}
\toprule
\textbf{System}	& \textbf{Dilute 2DEG}	& \textbf{Mott Organics}& \textbf{TMD Moir\'e Bilayers}\\
\midrule
  Transition Type &  continuous? 
            &  \hspace{2pt}weakly first order \newline (at $T < T_c \sim 0.01T_F$)
            &  continuous? \\
\midrule
  $\Delta$   & $|n-n_c|$ %
            & \hspace{14pt}$|P - P_c|^{\nu z}$,\newline$\nu z\approx 0.7 -1$ 
            & \hspace{14pt}$|E - E_c|^{\nu z}$,\newline$\nu z\approx 0.6$ \\
\midrule
 $\dfrac{1}{m^*}$      & $ |n-n_c|  $ 
            & ?    
            & ?\\\midrule
 $T_o$ &  \hspace{20pt}$|n - n_c|^{\nu z}$,\newline $\nu z\approx 1.6$  
 & \hspace{16pt}$|P - P_c (T)|^{\nu z}$,\newline$\nu z\approx 0.5 -0.7$
            & \hspace{16pt}$|E - E_c|^{\nu z}$,\newline$\nu z\approx 0.7$\\        
\midrule
  $T_{\mathrm{FL}}$   & ?  
            & $|P - P_c|$  
            &  \hspace{16pt}$|E - E_c|^{\nu z}$,\newline$\nu z\approx 0.7$  \\ \midrule
 $T_{max}$  & $|n-n_c|$
            & $|P - P_c|$  
            &  \hspace{16pt}$|E - E_c|^{\nu z}$,\newline$\nu z\approx 0.7$ \\

\bottomrule
\end{tabularx}
\end{table}

\subsection{Dilute 2DEG in Semiconductors}

In dilute two-dimensional electron gases (2DEG)~\cite{Fowler:1982}, the electron density can be quantified by the dimensionless parameter $r_s = 1/\sqrt{\pi n} a_B$ where $n$ is the electron density and $a_B$ the Bohr radius. The ratio of interaction energy versus kinetic energy scales as $r_s$, and therefore at large enough $r_s$ (of the order $r_s \sim 40$ in 2D) the electrons will spontaneously crystallize into a Wigner solid. In a two-dimensional Wigner crystal, the electrons form a triangular lattice with exactly one electron per unit cell---essentially forming a frustrated Mott insulator. When the electron density $n$ is varied, the size of the unit cell changes accordingly so that the Wigner crystal always remains fixed at one electron per unit cell.
The transition from an insulating Wigner crystal to a metal can therefore be plausibly viewed as a bandwidth-tuned Mott transition. Note that this is counter-intuitive: after all, one tunes the electron density! However, what matters is the electron density {\em counted per unit cell} and that remains constant. This idea suggests~\cite{camjayi2008coulomb,amaricci2010extended,Radonjic:2012gm} that the melting of a Wigner solid by increasing density should be viewed as a Wigner--Mott transition, possibly bearing many similarities to Mott transitions in narrow-band crystalline solids such as Mott organics or transition-metal oxides. If this viewpoint is correct, then the resulting metal above the transition should display resemble other strongly correlated Fermi liquids, a notion that is starting to gain acceptance on the base of recent experiments~\cite{shashkin2020manifestation, shashkin2022spin,moon2020quantum}.  

\textls[-15]{Experimentally, high-quality 2DEGs can be realized in  metal-oxide-semiconductor field-effect devices (MOSFETs) in various semiconductors~\cite{Kravchenko:1995PRB,Popovic:1997mit,shashkin2020manifestation,moon2020quantum}. Through electrostatic gating the electronic density can be elegantly tuned, typically in the range of \mbox{$n \sim 10^{10}$--$10^{12}$ cm$^{-2}$}. The peak electron mobility in ultra-clean samples can be as high as $10^4$ cm$^2$/V s~\cite{kravchenko2019}, which implies that down to very low temperatures the transport properties are dominated by electron-electron interactions (like Wigner crystallization) rather than extrinsic disorder effects. Lower-mobility devices have also been extensively studied (for a review see Chapter~5 of Ref.~\cite{kravchenko2017strongly}), displaying different types of metal-insulator transitions displaying electron glass dynamics~\cite{popovic2002glassy}, which we will not discuss here.}



\begin{figure}[H]

\includegraphics[width=4.6in]{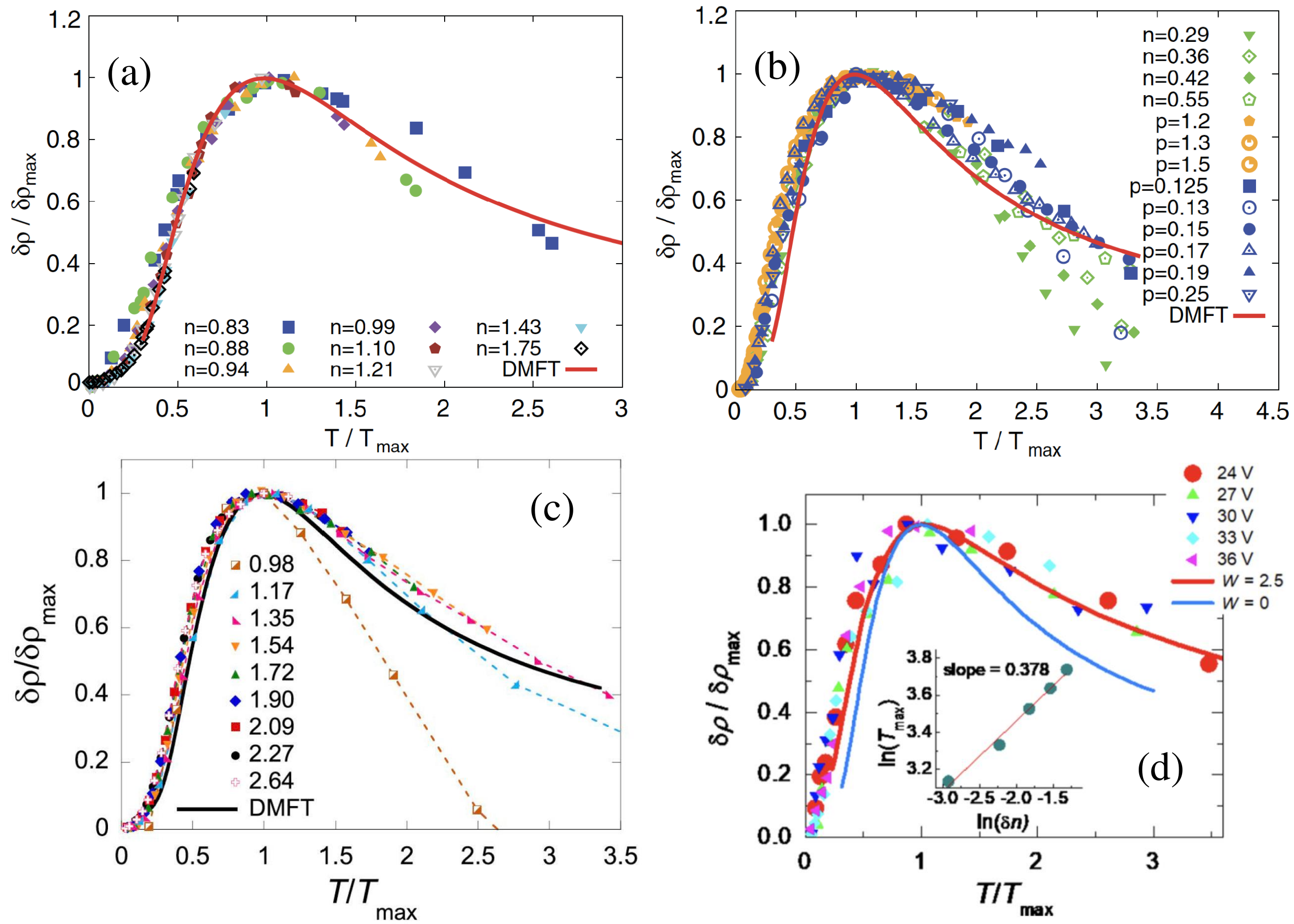}
\caption{\textls[-25]{Characteristic scaling of the resistivity maxima has been reported in several 2DEG electron systems in semiconductors: (\textbf{a}) Si-MOSFETs
~(adapted with permission from Ref. \cite{Radonjic:2012gm} Copyright 2012 American Physical Society); (\textbf{b}) p-GaAs/AlGaAs quantum wells~(adapted with permission from Ref. \cite{Radonjic:2012gm} Copyright 2012 American Physical Society); (\textbf{c}) SiGe/Si/SiGe quantum wells~(adapted with permisssion from Ref. \cite{shashkin2020manifestation} Copyright 2020 American Physical Society); (\textbf{d}) few layered-$MoS_2$~(Adapted with permission from Ref. \cite{moon2020quantum} Copyright 2020 American Physical Society). All data collapse to the same (theoretical) scaling function~\cite{Radonjic:2012gm} obtained from the Hubbard model at half-filling, in the vicinity of the Mott point.}
}
\label{Fig:Resistivitymaxima}
\end{figure}

Indeed, tuning the electronic density leads to insulating transport below a critical density, typically around $n_c \sim 10^{11}$ cm$^{-2}$~\cite{Kravchenko:1995PRB,Popovic:1997mit}, see Figure~\ref{Fig:Resistivity}a. Activated behavior is often observed close to the transition~\cite{shaskin2001prl,popovic2002glassy}, with the activation energy $\Delta \sim |n - n_c|$. Further on the insulating side disorder effects may become important, where Efros-Shklovskii hopping (and other effects of disorder) is also observed~\cite{Kravchenko:1995PRB}, but only at the lowest temperatures. 
On the metallic side, a pronounced resistivity drop (often by a factor of 10 or more) is observed~\cite{Kravchenko:1995PRB,shashkin2020manifestation} below the temperature $T_{max}\sim |n - n_c|$ which decreases as the transition is approached. Characteristic scaling of the resistivity maxima has been reported in several systems~\cite{shashkin2020manifestation, shashkin2022spin,moon2020quantum}, see Figure~\ref{Fig:Resistivitymaxima}, which has been interpreted as evidence for strong correlation effects. However, the expected $T^2$ dependence of the resistivity has not been observed, despite the reported  effective mass enhancement $(m^*)^{-1} \sim |n - n_c|$~\cite{kravchenko2002mass}, characteristic of correlated Fermi liquids. Quantum critical scaling collapse of the resistivity curves has also been demonstrated~\cite{Kravchenko:1995PRB} around the critical density, albeit excluding the lowest temperatures data, as shown in {Figure
}~\ref{Fig:QCscaling}a. This is achieved by rescaling $T$ by a crossover scale $T_o \sim |n - n_c|^{\nu z}$, with $\nu z\approx 1.6$. The resulting scaling function reveals surprising ``mirror symmetry''~\cite{kravchenko1997mirror}, which was phenomenological interpreted~\cite{gang41997mirror} as evidence ``strong-coupling quantum criticality''. Similar systems to these 2DEGs include the observation of a Wigner crystal in low-density doped monolayer WSe$_2$~\cite{Smolenski:2021hi}, where more detailed experiments still need to be performed. 

\subsection{Organic Compounds}

An organic compound~\cite{dressel2020advances} refers to a crystalline system where each unit cell contains an entire {\em molecule}, rather than just loosely bound ions. A particularly interesting class of organic compounds is based on the molecule bis-(ethylendithio)-tetrathiafulvalen (BEDT-TTF or ET), which can be fabricated with other ions into quasi-two-dimensional layered systems. Compounds based on BEDT-TTF exhibit a spectrum of interesting quantum matter phenomena, ranging from superconductivity~\cite{Jerome:1991jf} to electron glass behavior~\cite{Kagawa:2013hz}. 

Our interest goes out especially to $\kappa$-(ET)$_2$Cu[N(CN)$_2$]Cl and $\kappa\text{-(ET)}_2\text{Cu}_2\text{(CN)}_3$, where the molecules are organized in triangular lattice layers~\cite{dressel2020advances}. These materials are strongly correlated, and indeed, despite being half-filled they are insulating at ambient pressures. Due to the geometric frustration of the triangular lattice~\cite{Balents:2010ds}, no magnetic order has been observed in $\kappa$-Cu$_2$(CN)$_3$ and antiferromagnetic order only at relatively low temperatures ($T < T_N \approx 20K$) in $\kappa$-Cl. The absence of magnetic order is the strongest indication that $\kappa$-Cu$_2$(CN)$_3$ might realize a spin liquid ground state~\cite{kanoda2005spinliquid,pustogow2018natmat}.




Upon applying pressure, a zero-temperature {\em first-order} phase transition brings the system into a paramagnetic metallic phase at $p_c = 122$ MPa ($\kappa$-Cu$_2$(CN)$_3$) or $p_c = 24.8$~MPa ($\kappa$-Cl), see Figure~\ref{Fig:Resistivity}a~\cite{Furukawa:2015eo,Pustogow:2021npj}. The first order phase boundary ends in a critical point at $T_c = 20$ K or $T_c = 38$ K, respectively. It is important to emphasize that these temperatures are very small compared to the electronic energy scales. The Hubbard repulsion $U$ and bandwidth $W$ are both on the order of a fraction of eV~\cite{pustogow2018natmat}, which implies $T_c \ll U,W$. As such, even though the observed Mott criticality appears at nonzero temperatures, much of the observed phenomena above $T_c$ can be described {\em as if} the system resides in the vicinity of a quantum critical point~\cite{Furukawa:2015eo}.

\begin{figure}[H]

\includegraphics[width=5in]{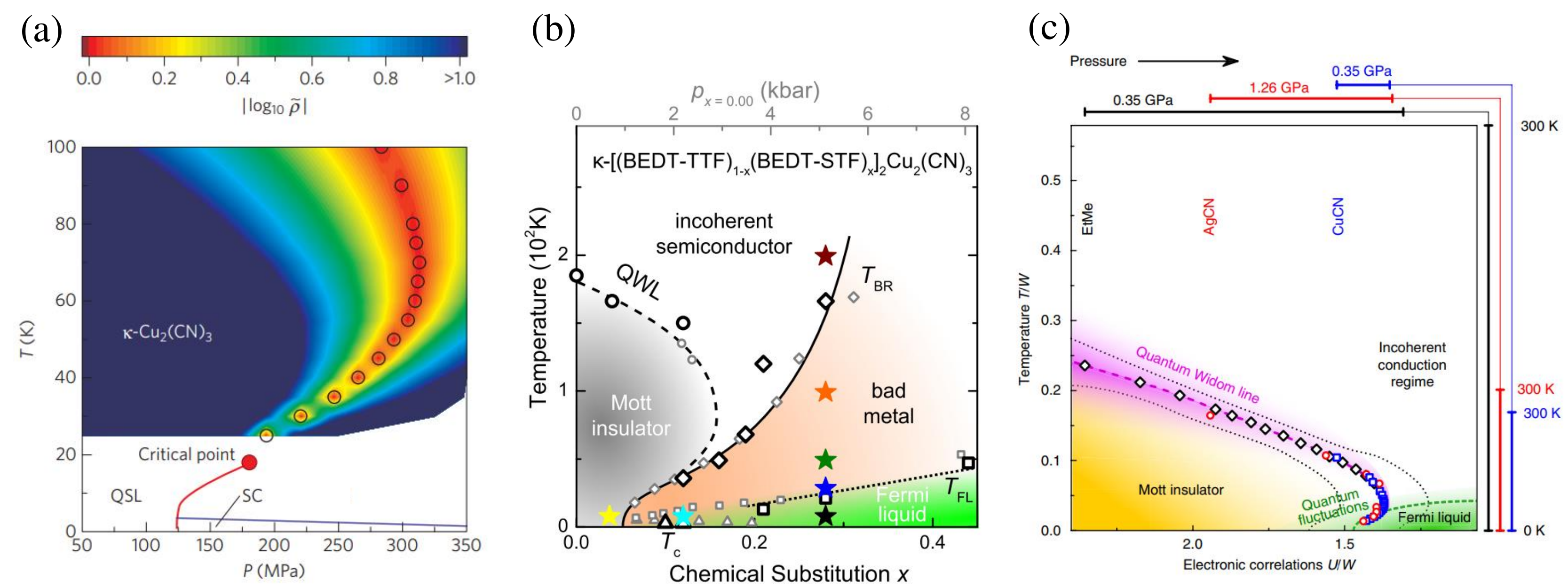}
\caption{\textls[-15]{Finite temperature phase diagram of the Mott organic materials. (\textbf{a}) first-order phase transition line, as observed in
 $\kappa$-Cu$_2$(CN)$_3$ (adapted with permission from Ref.\cite{Furukawa:2015eo} Copyright 2015 Springer Nature) at $T < T_c \sim 20K$, displaying ``Pomeranchuk'' behavior~\cite{pomeranchuk1950jetp}, by ``sloping'' towards the metallic phase. The corresponding ``Quantum Widom Line''~\cite{vucicevic2013finite} arises at $T > T_c$, which is identified as the center of the quantum critical region~\cite{Terletska2011} with resistivity scaling~\cite{Furukawa:2015eo}. (\textbf{b})~Phase diagram~\cite{Pustogow2021FL} for  $\kappa$-Cu$_2$(CN)$_3$ over a broader $T$-range, displaying the convergence of the quantum Widom line (QWL) on the insulating side, and the ``Brinkman-Rice'' line ($T_{BR} = T_{max}$, which intersect at the critical end-point $T = T_c$. The Fermi-Liquid line $T_{FL} < T_{BR}$ is also shown.
(\textbf{c}) The universal phase diagram for a series of spin-liquid Mott organics compounds was established~\cite{pustogow2018natmat} by rescaling the temperature $T$ and the interaction strength $U$ by the respective electronic bandwidth $W$. The parameters $W$ and $U$ were independently measured~\cite{pustogow2018natmat} for each material using optical~conductivity.}}
\label{Fig:QCphasediagram}
\end{figure}

\begin{figure}[H]
\includegraphics[width=0.95\textwidth]{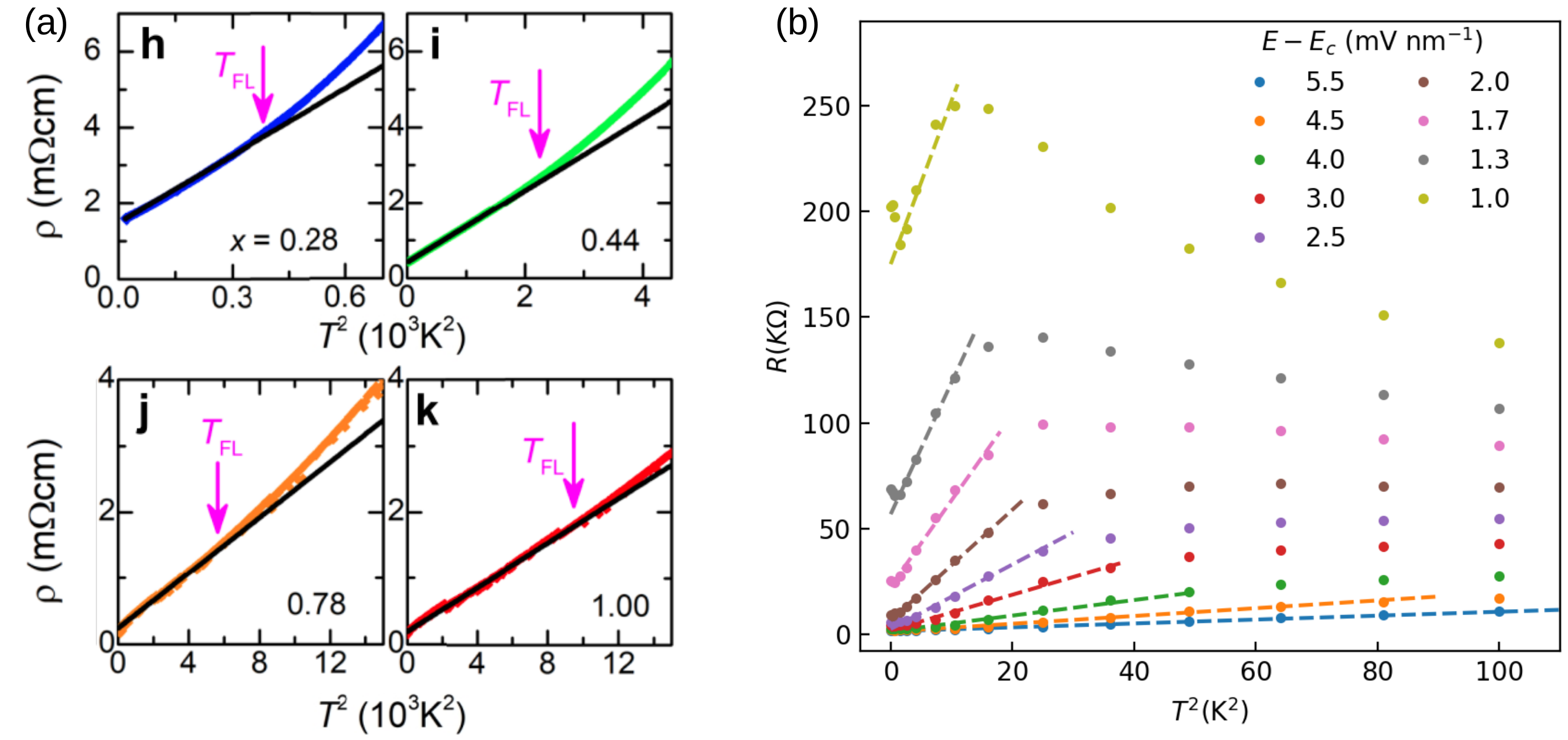}
\caption{Fermi liquid behaviour at low temperatures, for (\textbf{a}) Mott organic material $\kappa-[\text{(BEDT-TTF)}_{1-x}\text{(BEDT-STF)}_x]_2\text{Cu}_2\text{(CN)}_3$~\cite{Pustogow2021FL} and
 (\textbf{b}) MoTe$_2$/WSe$_2$ moir\'{e} bilayers (Data imported from~\cite{Li2021ContinuousMT}). Clear $\rho = \rho_0 + AT^2$ behavior is observed in both cases, up to a temperature scale $T_{FL}$ that seems to decrease linearly towards the metal-insulator transition. The resistivity curves can be collapsed by plotting $\rho(E,T)/\rho_c(T)$ vs. $T/T_0$ where $T_0 \sim|E-E_c|^{0.70\pm0.05}$, see Figure~\ref{Fig:QCscaling}c. Note that this crossover scale seems to follow both the gap size on the insulator, as well as the destruction of the Fermi liquid on the metallic side.
}
\label{Fig:FermiLiquid}
\end{figure}

\textls[-30]{Above $T_c$, a crossover pressure $P_c(T)$ can traced ~\cite{Furukawa:2015eo} where the measured resistivity exhibits an inflection point, see Figure~\ref{Fig:QCphasediagram}. This defines the {\em ``quantum Widom line''} (QWL)~\cite{vucicevic2013finite} by analogy to the standard liquid-gas crossover. Defining the critical resistivity $\rho_c(T)$ to be the resistivity along the QWL, all resistivity curves collapse onto each other when plotted as $\rho(P,T)/\rho_c(T)$ vs. $T/T_0(P)$, as shown in {Figure
}~\ref{Fig:QCscaling}b. Here the scale $T_o(P)$ reflects a critical energy scale that vanishes at the critical pressure, $T_o \sim |P-P_c|^{0.6}$ for Cu and $T_o \sim |p-p_c|^{0.5}$ for $\kappa$-Cl ~\cite{Furukawa:2015eo}. On the insulating side of the transition, the resistivity is approximately activated $\rho \sim \exp (\Delta/T)$~\cite{pustogow2018natmat}. On the metallic side, it follows~\cite{Pustogow2021FL} the standard Fermi liquid behavior at low temperatures $\rho (T) = \rho_0 + A T^2$, up to a temperature scale  $T_{FL}$, see Figure~\ref{Fig:FermiLiquid}a. This destruction of the Fermi liquid seems to correspond to the appearance of a maximum in the resistivity~\cite{Pustogow:2021npj}. }

\begin{figure}[H]
\includegraphics[width=2.5in]{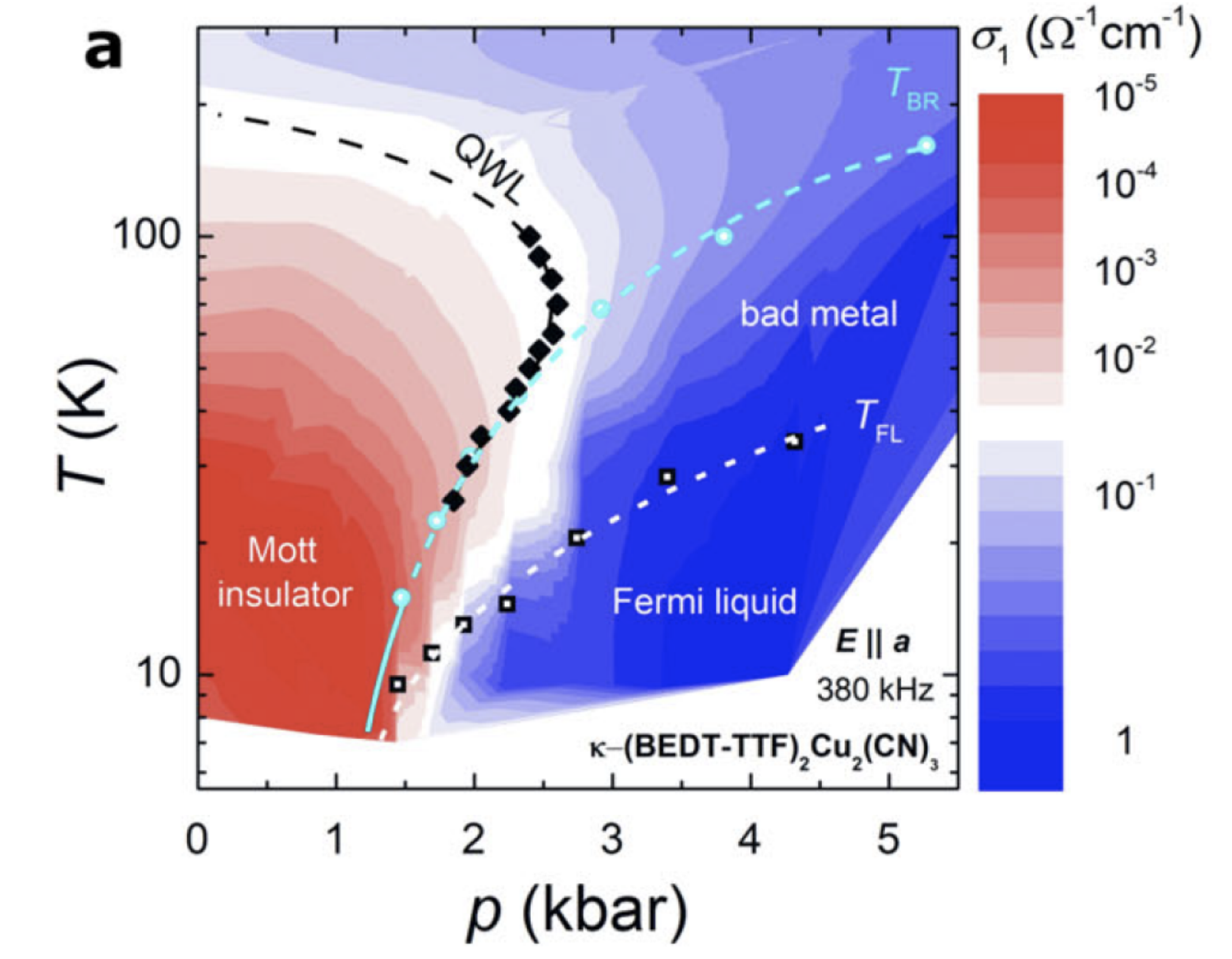}
\includegraphics[width=2.3in]{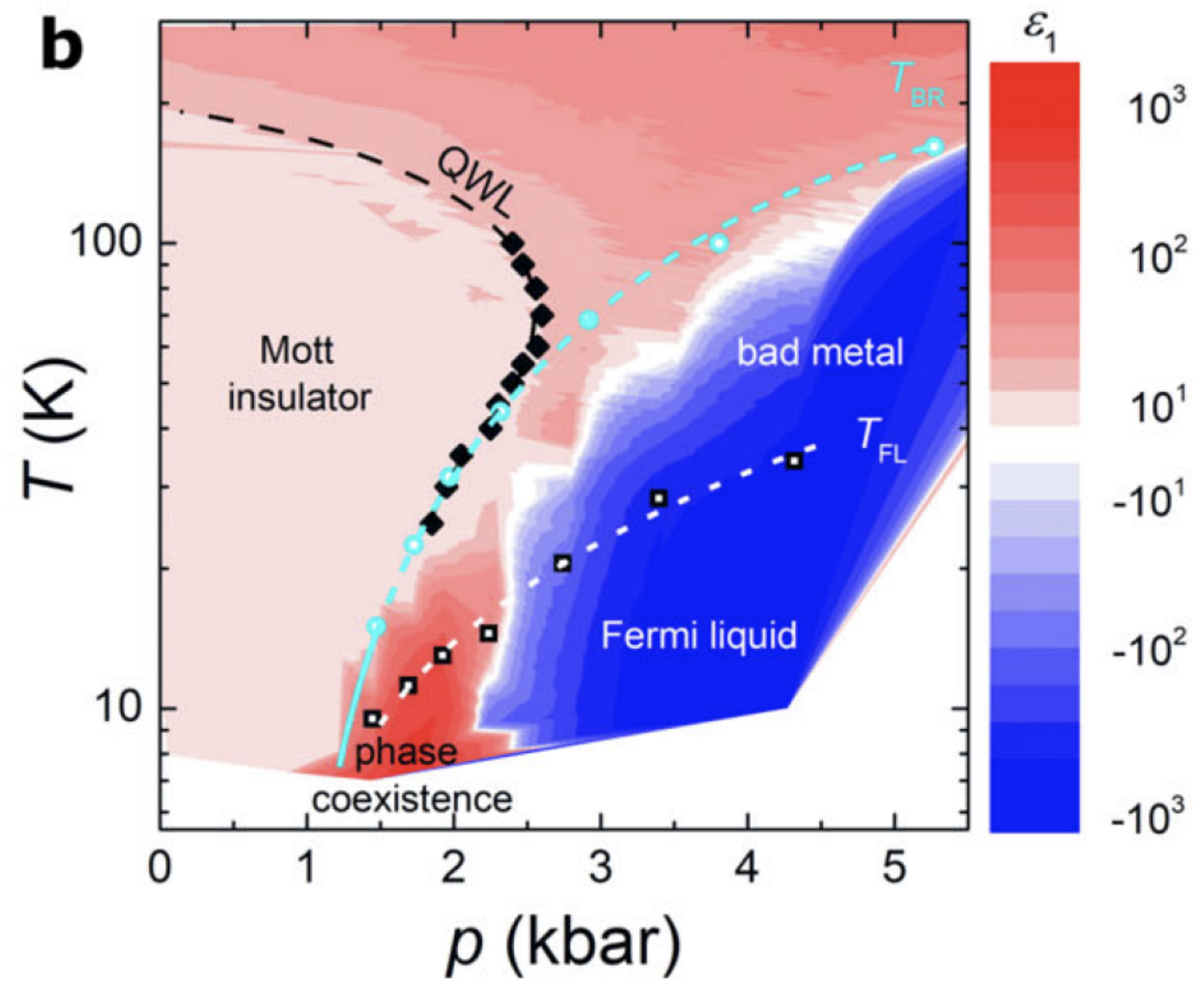}
\caption{Transport behavior vs. dielectric response across the phase diagram of
 $\kappa$-Cu$_2$(CN)$_3$~\cite{Pustogow:2021npj}. (\textbf{a})~DC transport shows only very gradual change across the BR line (resistivity maxima), and cannot one see any clear indication of the phase coexistence region. (\textbf{b}) In dramatic contrast, the low-frequency dielectric function $\epsilon_1$ assumes small positive values in the Mott insulator (pale pink), and large negative values in the quasiparticle regime (deep blue); we clearly see the boundaries of these regimes tracing the QWL and the BR line (following $T_{max}$), as observed in transport. Remarkably, ``resilient'' quasiparticles~\cite{kotliar2013resilient} persist past the Fermi Liquid line, at $ T_{FL} < T < T_{BR} = T_{Max}$, where bad metal behavior~\cite{kivelson1992badmetal} (metallic transport above the Mott-Ioffe-Regel~\cite{hussey2004mirlimit} limit is observed). At low temperature, the Mott point is buried below the phase coexistence dome, which is vividly visualized through colossal dielectric response ($\epsilon_1 \sim 10^3$--$10^4$).}
\label{Fig:DielectricOrganics}
\end{figure}

\textls[-15]{In addition to transport measurements, and in contrast to other systems we consider, Mott organics have also been carefully investigated using optical probes. This allowed to directly identify~\cite{pustogow2018natmat} the quantum Widom line, which is back-bending towards the insulating side at higher temperatures following the closing of the Mott gap. In addition, the ``Brinkman--Rice'' line traced by $T_{max}$ was identified as marking the thermal destruction of Landau quasiparticles, as seen by the vanishing of the Drude peak in the optical conductivity~\cite{Pustogow2021FL}, see Figure~\ref{Fig:DielectricOrganics}. }

Finally, the controversy about the presence or absence of the low-$T$ phase coexistence region has been resolved in Mott organics, by using dielectric spectroscopy~\cite{Pustogow:2021npj}. Its precise location on the phase diagram has been identified by the observation~\cite{Pustogow:2021npj} of colossal dielectric response, as a smoking gun for percolative phase coexistence. In addition, the same technique was able to demonstrate the coincidence of the resistivity maxima in the (uniform) metallic phase, with the thermal destruction of Landau quasiparticles. This is seen as a dramatic drop and a change~\cite{Pustogow:2021npj} of sign of the dielectric function at $T < T_{BR} = T_{Max}$. These experimental results are shown in Figure~\ref{Fig:DielectricOrganics}, and discussed in more detail in Section~\ref{Sec:Dielectric}.


\subsection{Moir\'{e} Materials}

\textls[-25]{The most recent addition to the field of strongly correlated systems are {\em moir\'{e} materials}. These are bilayer structures made of Van der Waals materials such as graphene and transition-metal dichalcogenides (TMDs). A lattice mismatch or relative twist angle between the layers causes a large-scale geometric ``moir\'{e}'' pattern. This larger unit cell (typically in the range of 5--10 nm) drastically reduces the effective electron kinetic energy such that the bandwidth is on the order of $W\sim10$ meV. As a result, the systems become strongly correlated, with $U/W \sim 10$ or larger. Using electrostatic gating one can tune the electronic density, typically in the range of a few electrons or hole per moir\'{e} unit cell (corresponding to $n \sim 10^{12}$~cm$^{-2}$).}

While the most famous of correlated moir\'{e} materials is without a doubt twisted bilayer graphene, more convincing evidence for Mott correlations has so far only been observed in TMD bilayers. Here, we focus on one particular system: the heterobilayer MoTe$_2$/WSe$_2$~\cite{Li2021ContinuousMT}. The lattice mismatch  between MoTe$_2$ and WSe$_2$ gives rise to a moir\'{e} period of $a_M \sim 5$ nm. At half-filling of the first valence band, an insulating phase appears which can be tuned into a metal by applying a vertical displacement field $E$. This flat valence band can be described by a spin-orbit coupled triangular lattice Hubbard model, where the displacement field $E$ tunes the bandwidth~\cite{Rademaker2021TMD}. 

The temperature-dependent resistivity across the transition is shown in Figure~\ref{Fig:Resistivity}c. On the insulating side, the system has well-defined activated behavior of the resistivity with a gap $\Delta$ continuously vanishes as the critical displacement-field value is approached, $\Delta \sim |E-E_c|^{0.60\pm0.05}$. At the critical point, the resistivity is claimed to follow a powerlaw, $\rho_c \sim T^{-1.2}$, although the reliability of the low-$T$ data may be questionable. On the metallic side, the low-$T$ resistivity follows the Fermi liquid law $\rho(T) = \rho_0 + AT^2$ where the quadratic prefactor diverges $A \sim |E-E_c|^{-2.8\pm 0.2}$. This clear Fermi liquid does not persist up to all temperatures, instead, a resistivity maximum appears at $T_{\mathrm{max}} \sim |E-E_c|^{0.70\pm0.05}$, see Figure~\ref{Fig:FermiLiquid}b. No magnetic order has been observed, which might be due to the geometric frustration of the triangular moir\'{e} lattice structure. Remarkably, this experiments uses the recently-developed {\em excitonic sensor}~\cite{Li2021ContinuousMT} , which allows the measurements of the spin susceptibility across the transition. This reveals Curie-law behavior over a broad temperature range, thus demonstrating the presence of localized magnetic moments, as expected for a Mott system. 

Finally, among other moir\'{e} systems it is worth mentioning the twisted hetero-bilayer WSe$_2$~\cite{Ghiotto:2021qc}. Though it was claimed to exhibit some sort of Mott-related criticality, it is not certain whether a true insulating phase has indeed been observed given that the activated transport does not continue down to the lowest measured temperatures. In addition, insulating behavior seems to disappear above a relatively low  of the order of $ T^* \sim$ \mbox{5--10 K,} which is similar to twisted bilayer graphene, but much smaller than the estimated bandwidth of the order of $W \sim$ 100--200 K.  Furthermore, no clear resistivity maxima on the metallic side have been observed. Whether or not tbWSe$_2$ can be classified as a true Mott insulator is therefore quite controversial. Alternatively, the observed behavior could be a result of some sort of magnetic order, which may arise close to half filling even in weakly-coupled systems. 


\subsection{Universal Criticality}

The resistivity curves of the three systems, as shown Figures~\ref{Fig:Resistivity} and \ref{Fig:QCscaling}, show remarkable universality,  reflected in the fact that all curves can be collapsed by scaling with $T_0 \sim |\delta|^{z\nu}$ where $\delta$ is the tuning parameter and $z\nu$ the critical exponent. It is important to realize, however, that the precise scaling procedure applied  was not identical in the three cases, and the resulting critical exponent also somewhat depend on the system.

So what is {\em different} between these systems? Let us first focus on the energy scale. The typical bandwidth $W$ ranges from $\sim$100 s meV in Mott organics, to $\sim$ 10 s meV in the moir\'{e} systems, to 0.1--1 meV in the dilute 2DEGs. A finite temperature critical point is only observed in the organics, though at about $T_c \sim 1\% W$---which leaves open the possibility that a finite $T$ critical endpoint exists in the other two systems. Indeed, most experiments are performed (so far) above the Kelvin range in moir\'e materials, and above 100 mK in 2DEG, which makes it hardly possible to reliably explore the $T$-range below few percent of the~bandwidth.

Secondly, in order to achieve quantum critical scaling in Mott organics, one needs to first identify a Widom line as a demarcation of the finite-temperature crossover from insulator to metal. A similar analysis was carried out for moir\'e materials, although the obtained Widom line displayed no apparent ``curving'' as a function of temperature. This is manifestly not performed 
in the dilute 2DEGs. It might be interesting to see whether better collapse can be achieved through such a method.

Thirdly, with a bit of good-will, the critical exponents in organics and Moir\'{e} systems are in the same ballpark; whereas the critical exponent in the dilute 2DEGs with $z\nu = 1.60$ is significantly larger. It is also important to realize that in 2DEGs there has not been a clear observation of a Fermi liquid regime---unlike in organics and Moir\'{e} systems, see Figure~\ref{Fig:FermiLiquid}. These ways in which 2DEGs stand out might be related to the fact that there is no underlying (Wigner) lattice on the metallic side, which could point to a perhaps nontrivial role of significant charge density fluctuations on the metallic side, an effect not present in lattice Mott systems. 

\section{Competing Theoretical Pictures} 
\label{Sec:Theory}

As we mentioned in Section~\ref{Sec:IntroMott}, the observation of Mott criticality and scaling opens big questions on the theoretical front. Currently, there exist {\em three} main different physical pictures to address these issues.

A true Mott transition should not be hidden by some period doubling symmetry breaking. The Lieb-Schultz-Mattis theorem states that in absence of spin order, the ground state of Mott insulator must be a spin liquid~\cite{Savary:2016fk}. This leads directly to the first theoretical picture: Mott criticality can only occur if the Mott phase is a spin liquid, where inter-site spin correlations play an important role. The theory of Senthil~\cite{Senthil:2008ki} chooses this path, by introducing an explicit spinon theory of the Mott spin liquid.

Alternatively, one focuses on local electronic processes only, ignoring inter-site spin correlations. Then the Mott transition at low temperature becomes first-order; however, it is only {\em weakly} first order. A first order transition line always ends at a critical point $T_c$, and as long as $T_c$ is sufficiently low compared to any experimental scale, one still finds criticality and scaling. This is the picture emerging from Dynamical Mean Field Theory (DMFT)~\cite{Georges:1996un}, a strong-coupling self-consistent approach to calculate the local electronic self-energy.

The third picture again accepts the first-order nature of a Mott transition, but this time embraces it. A first-order transition is always accompanied by a region where both phases coexist. Minor disorder or self-generated pattern formation~\cite{SpivakKivelson2004PhRvB,Spivak2006} can smear this phase coexistence region into a continuous-looking transition exhibiting nontrivial electron dynamics. This is the `percolation theory' picture of Mott criticality.

The goal of this review paper is to put the main theoretical predictions next to the experimental findings. As such, we will not dive into the pros and cons of each theoretical picture. A summary of the main theoretical predictions is provided in Table~\ref{Tab:Theory}.

\begin{table}[H]
\caption{A summary of predictions from competing theoretical pictures. The expected transition type differs between the three pictures, with observable differences in the behavior of 
the mass enhancement $m^*$, the Kadowaki--Woods ratio $A/(m^*)^2$, the destruction of the Fermi liquid at $T_{\mathrm{FL}}$, and the appearance of a resistivity maxima at $T_{max}$. Details are provided in the text below.\label{Tab:Theory}}
\newcolumntype{C}{>{\centering\arraybackslash}X}
\begin{tabularx}{\textwidth}{CCCC}
\toprule
 \textbf{Theory Predictions} & \textbf{2D Spinon Theory} &\textbf{DMFT} & \textbf{Percolation Theory}\\
\midrule
  Transition Type    &  continuous 
            & weakly first order \newline(at $T < T_c \sim 0.01T_F$)
            & first order \\
\midrule
  $\Delta$   & \hspace{14pt}$|g-g_c|S^{\nu z}$, \newline $\nu z= 0.67$
            & \hspace{12pt}$ |U - U_{c1}|^{\nu z}$,\newline$\nu z\approx 0.8$ 
            & remains finite \\
\midrule
 
 $m^*$      &  weak: $\ln \frac{1}{|g-g_c|} $ 
            & strong: $|U - U_{c2}|^{-1}$    
            & no divergence \\\midrule
 $A/(m^*)^2$&  ?
            & \hspace{12pt}constant \newline(KW law obeyed)
            & diverges:  $(x_o - x_c)^{-t}$; $t = s/m$\\        
  \midrule
  $T_{\mathrm{FL}}$   & $|g - g_c|^{2\nu} $ 
            & $|U - U_{c2}|$ 
            &  $T^*\sim |x_o - x_c|$\\ \midrule
 $T_{max}$  & $T_{max} = \infty$
            & $|U - U_{c2}|$ 
            & $T^* \sim |x_o - x_c|$\\
\bottomrule
\end{tabularx}
\end{table}

\subsection{Spin Liquid Picture of the Mott Point}
\label{Sec:SpinonTheory}

A popular approach to describe a spin liquid state is through {\em spin-charge separation}. In Ref.~\cite{Senthil:2008ki}, the electron is split into a charge-0 spin-$1/2$ fermionic {\em spinon} $f$ and a charge-$e$ spin-0 bosonic {\em chargon} $b$. The Mott transition, in this picture, amounts to the condensation of the chargon field, whose critical behavior falls within the 3D $XY$ universality class. The Fermi liquid corresponds to the condensed phase of the chargon, whereas the Mott insulator corresponds to a gapped phase of the charged boson. The splitting of the electron leads to redundant degrees of freedom described by an emergent gauge field. Fluctuations of this gauge field lead to a logarithmic enhancement of the quasiparticle effective mass,
\begin{linenomath}
\begin{equation}
    m^* \sim \ln \frac{1}{| g - g_c|},
\end{equation}
\end{linenomath}
where $g$ is the tuning parameter and $g_c$ is the critical value. However, as in any theory with a non-local electronic self-energy, the quasiparticle residue $Z$ is {\em not} simply proportional to the inverse effective mass; instead $Z \sim |g-g_c|^\beta / \ln  \frac{1}{| g - g_c|}$. Furthermore, approaching the Mott transition from the metallic side the spin susceptibility $\chi$ remains constant whereas the compressibility $\kappa$ vanishes. Physically, these effects result from important inter-site spin correlations, where a gapless spin liquid can be viewed as a certain superposition of spin singlets formed by pairs of spins in the Mott insulating state. As a result, there emerges a finite gap $\delta$ to charge excitations, while the rearrangement of singlets leads to characteristic gapless spin excitations with fermionic quasiparticles. This picture is a specific realization of the famous RVB picture of Baskaran and Anderson~\cite{BASKARAN1987973}, first proposed in the context of high-$T_c$ superconductors. 

Another significant consequence of describing the Mott transition as chargon condensation, is that the $T=0$ conductivity is not continuous. The electron resistivity will display a {\em universal} jump from a (disorder)-dependent constant value $\rho = \rho_0$ in the Fermi liquid; to $\rho = \rho_0 + \frac{Rh}{e^2}$ (with $R$ of order one) at the critical point; to $\rho = \infty$ in the Mott insulator. On the metallic side, the Fermi liquid is predicted to break down above $T_{\mathrm{FL}} \sim | g - g_c|^{2 \nu}$ and give rise to a {\em marginal} Fermi liquid state, which in turn survives up to $T_{\mathrm{MFL}} \sim |g - g_c| ^\nu$. In both cases, $\nu = 0.67$ is the 3D $XY$ correlation length exponent. On the insulating side, the boson condensation picture implies that the charge gap vanishes as $\Delta \sim | g- g_c|^\nu$. The spinons, however, remain gapless and form a spinon Fermi surface, with low-temperature specific heat scaling as $C \sim T^{2/3}$.



Note that the original work in Ref.~\cite{Senthil:2008ki} does not directly provide a detailed description for finite temperature dependence of the resistivity, and thus no explicit prediction for a possible deviation from the Kadowaki-Woods (KW) law ($A/(m^*)^2 \approx$ constant)~\cite{jacko2009unified}. On the other hand, Ref.~\cite{jacko2009unified} presents arguments that the physical requirement for the validity of the KW law is the locality of the electronic self-energy (as in DMFT theory)~\cite{Georges:1996un}, a condition which is not obeyed by the RVB-type spin-liquid theories such as the Senthil's spinon~picture.

\begin{figure}[H]
\includegraphics[width=0.8\textwidth]{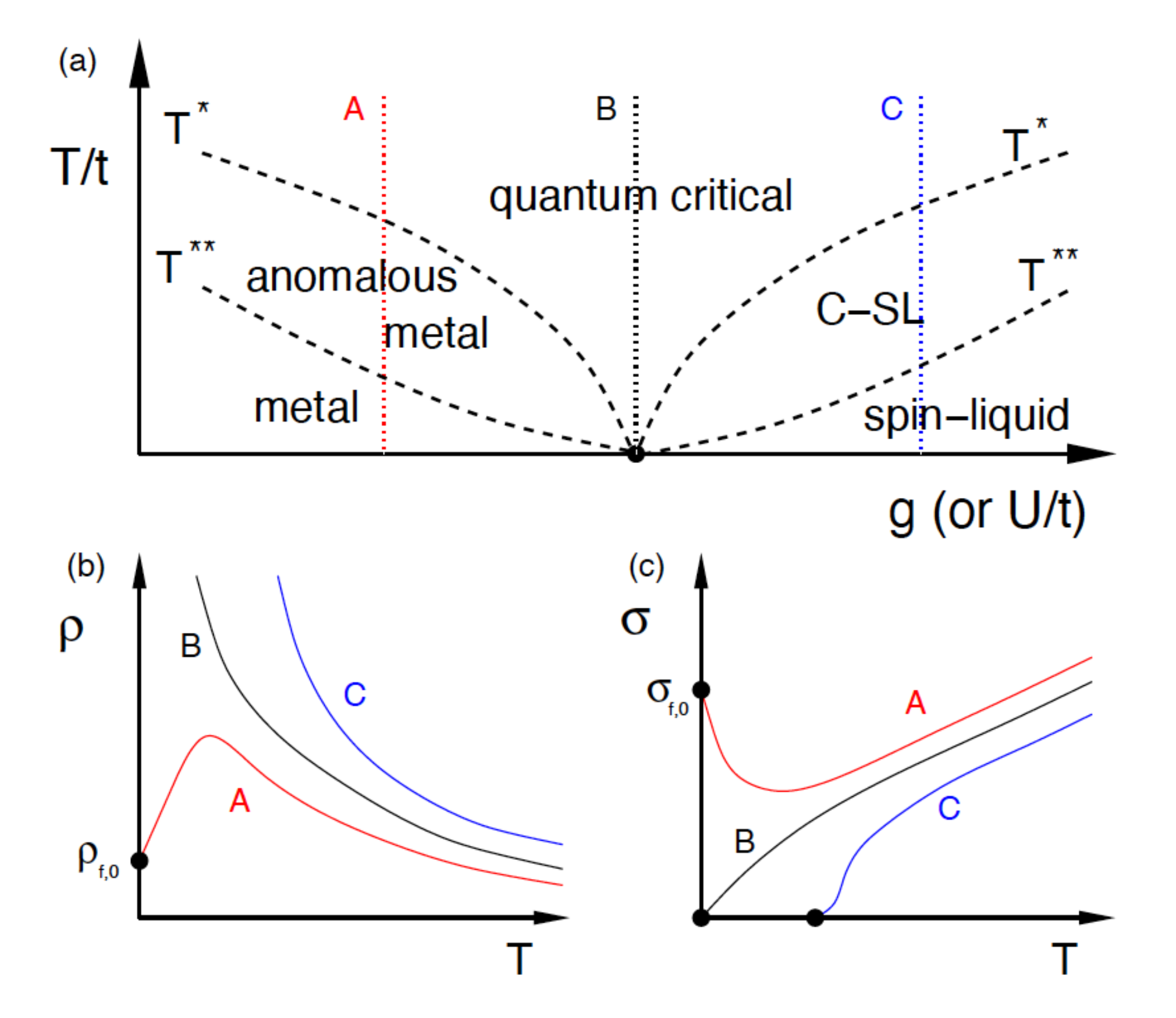}
\caption{Predictions of the spinon theory (reprinted with permission from Ref. \cite{senthil2009D3} Copyright 2009 American Physical Society).
(\textbf{a}) The phase diagram features a quantum critical point at $T=0$, and two distinct finite-$T$ crossover scales $T^*$ (above which the system is quantum critical) and $T^{**}$ (below which the system is either a metal or a gapless spin liquid). (\textbf{b}) and (\textbf{c}) Resistivity and conductivity along the lines A, B and C in the phase diagram in (\textbf{a}). Critical resistivity is predicted to diverge as $\rho_c(T) \sim 1/t$ in $d=3$, leading to resistivity maxima on the metallic side (coductivity minima). In contrast, the same theory predicts finite critical resistivity  $\rho_c(T) \sim \rho^*$ in $d=2$~\cite{Senthil:2008ki}, hence monotonic behavior on both sides of the transition and no resistivity maxima.} 
\label{Fig:Senthil}
\end{figure}

It should be stressed that the spinon theory makes one sharp prediction about finite temperature transport in the critical regime. Namely, the {\em critical resistivity curve}  is predicted to assume a universal power-law form $\rho_c(T) \sim 1/T$ in $d=3$~\cite{senthil2009D3}, but remain $T$-independent in $d=2$, see Figure~\ref{Fig:Senthil}. This therefore leads to distinct resistivity maxima in 3D, but {\em not} in 2D~\cite{Senthil:2008ki}, where monotonic $T$-dependence should be found on both sides of the transition, albeit with opposite slope. Physically, this difference reflects the proposed importance of ``infrared'' (IR, long distance) effects due to gauge fields, which should have strong (spatial) dimensionality dependence. Concerning quantum critical scaling, it is interesting that this theory proposes the emergence of {\em two} crossover temperature scales, both of which vanish at the transition.

We finally mention that a similar spin-charge separation theory has been very recently proposed to also describe the Wigner--Mott transition in TMD bilayers, where a possible role of charge fluctuations has also been discussed for the metallic side~\cite{XuJianXu2021arXiv,Musser:2021arXiv}.

\subsection{Dynamical Mean Field Theory Picture of the Mott Point}

Our second theory of interest is {\em Dynamical Mean Field Theory} (DMFT), which explicitly ignores all nonlocal (spin or charge) spatial correlations, and therefore aims to self-consistently calculate the local electronic self-energy $\Sigma(\omega)$~\cite{Georges:1996un}.  Physically, its real part describes the modifications of the electronic spectra, while its imaginary part encodes the frequency and temperature dependence of the electron-electron scattering rate. In this way, this theory is not limited to low-temperature excitations only, but is able to capture strong inelastic scattering at high temperatures, and therefore describe both the (coherent) Fermi liquid regime, and also the incoherent high-temperature transport, for example the famed bad metal behavior~\cite{kivelson1992badmetal,vuvcivcevic2015bad} above the MIR limit~\cite{hussey2004mirlimit}.

While there exist some limiting cases where an analytic solution is possible, it is mainly a numerical approach at finite temperature. DMFT is exact in the limit of large coordination, which physically corresponds to maximal magnetic frustration. Therefore, in the simplest implementation, DMFT describes Mott physics in absence of any magnetic order, nor does it include any (inter-site) spin liquid correlations. We should mention that extensions of DMFT have recently been proposed~\cite{lee2016fate} that include spinon effects, based on an alternative (matrix M,N) rotor representation. This theory, which includes some dynamical effects even at the saddle-point level, suggest that coherent spinon excitations are very fragile to charge fluctuations emerging upon the closing of the Mott gap, suppressing the spin liquid correlations not only on the metallic side, but also within the critical region. We will not further discuss these most sophisticated approaches here, but will limit our attention to the predictions of the simplest single-site DMFT theory. 

\begin{figure}[H]
\includegraphics[width=5.2in]{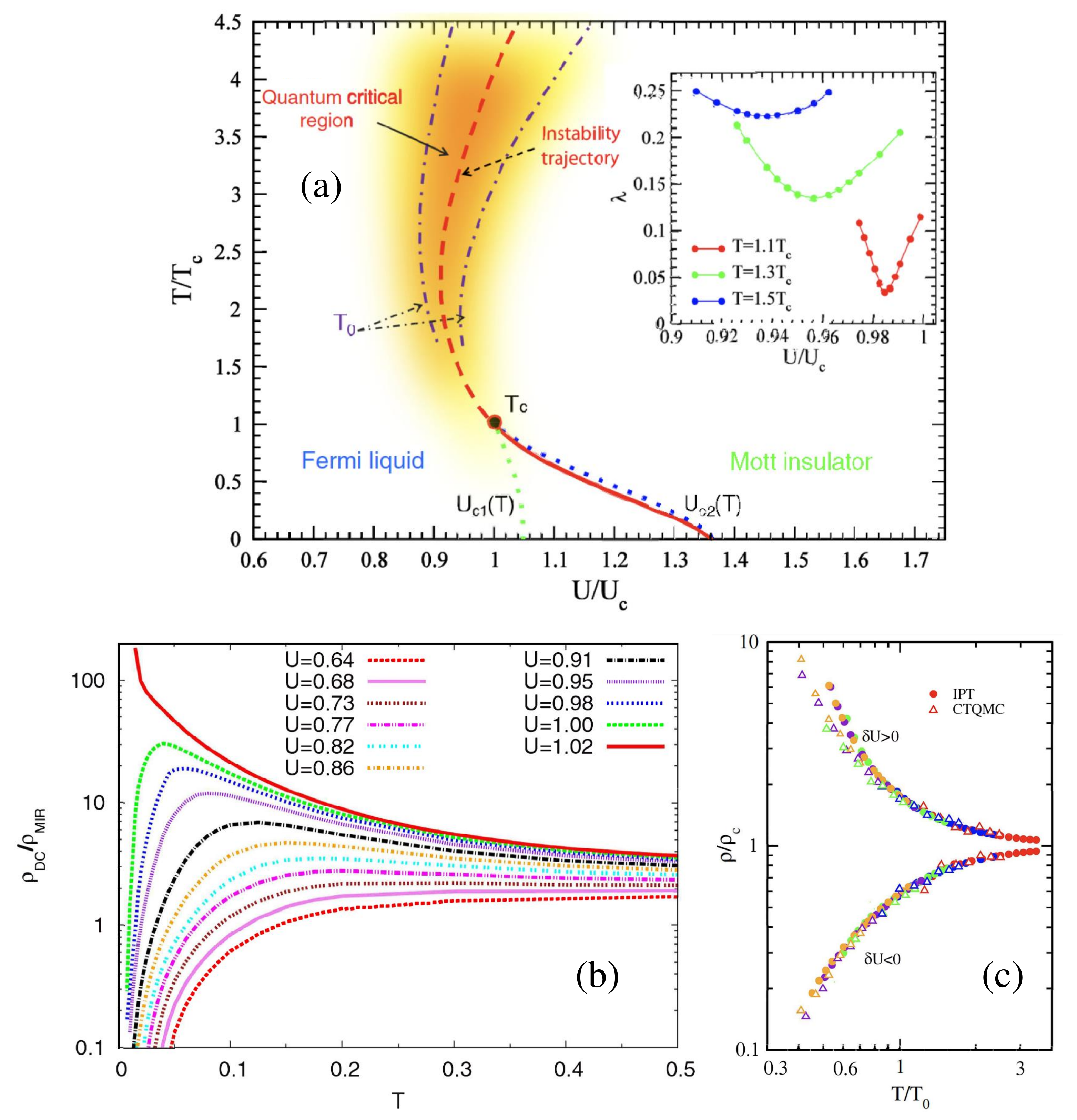}
\caption{Predictions of DMFT theory. (\textbf{a}) Phase diagram featuring a phase coexistence region at $T < T_c$, and a Quantum Critical region centered around the Quantum Widom Line (QWL) ~(adapted with permission from Ref. \cite{Terletska2011} Copyright 2011 American Physical Society). (\textbf{b}) Resistivity (normalized by the Mott-Ioffe-Regel (MIR) limit) as a function of temperature $T$ across the transition. Note the pronounced resistivity maxima on the metallic side ~(adapted with permission from Ref. \cite{Radonjic2010prb} Copyright 2010 American Physical Society. (\textbf{c}) scaling collapse of the resistivity curves, displaying pronounced ``mirror symmetry'' of the two branches (adapted with permission from Ref. \cite{Terletska2011} Copyright 2011 American Physical Society). }
\label{Fig:DMFT}
\end{figure}

When applied to the single-band Hubbard model on a frustrated lattice (such as the triangular lattice), DMFT predicts that on the metallic side the quasiparticle mass diverges linearly $m^* \sim | U-U_{c2}|^{-1}$ at a critical value $U_{c2}$ (similar to the prediction of the Brinkman--Rice (BR) theory of the Mott transition~\cite{BrinkmanRice1970}). The quasiparticle weight $Z$ is inversely proportional to $(m^*)$. Similarly, other features of the Fermi liquid such as the Kadowaki-Woods law $A \sim (m^*)^2$ are upheld. This Fermi liquid behavior persists up to a temperature $T_{\mathrm{FL}}$ that vanishes linearly when approaching $U_{c2}$. Interesting, at $T_{\mathrm{max}} \sim T_{\mathrm{FL}}$ the resistivity exhibits a maximum~\cite{Radonjic:2012gm}. On the insulating side, there exist no well-defined quasiparticles as the self-energy diverges, $\Sigma(\omega) \sim 1/\omega$. The electronic spectrum is split into an upper and lower Hubbard band, separated by a gap that remains nonzero at $U_{c2}$. The insulating state becomes unstable at a lower value of the interaction $U_{c1} <U_{c2}$, where the gap closes $\delta \sim |U - U_{c1}|$, $^{\nu z}$, $\nu z\approx 0.8$~\cite{vojta2019prb}. As a result, there emerges a  low-$T$ first-order metal-insulator transition, and an associated phase coexistence region at $U_{c1} < U < U_{c2}$~\cite{Georges:1996un}. These main predictions are summarized in Figure~\ref{Fig:DMFT}. 

At nonzero temperature, the first-order transition line ends at a critical point at a temperature $T_c \approx 0.015 W$, significantly smaller than the bare bandwidth $W$. At temperatures $T \gg T_c$ the results can be viewed as effectively quantum critical~\cite{Terletska2011,vojta2019prb}.  This quantum critical regime is centered around the so-called quantum Widom line (QWL)~\cite{vucicevic2013finite}, which physically represents a finite-temperature instability trajectory of the insulating phase, as shown in Figure~\ref{Fig:DMFT}a. It extends the first-order line past $T = T_c$, and can experimentally be detected from an inflection point analysis~\cite{Furukawa:2015eo} of the resistivity curves. In this regime, the resistivity satisfies the scaling law $\rho(T, \delta U) = \rho_c(T) f(T/T_0(\delta U))$ with a crossover temperature scale $T_o \sim |\delta U|^{ \nu z}$ where $ \nu z \approx 0.6$, see Figure~\ref{Fig:DMFT}c. The crossover scale $T_o$ is a property of the quantum critical regime and should not be confused with the low-temperature scaling of the Fermi liquid temperature $T_{FL}$. DMFT therefore predicts two different regimes of scaling: the quantum critical regime at $T \gg T_c$, and the metal regime dominated by scaling in $T_{\mathrm{FL}}$.

\subsection{Percolative Phase Coexistence Picture}

In the early theories of both Mott insulators~\cite{mott1990metal} and Wigner crystals, the transition from insulator to metal was often assumed to be robustly first order, at least at sufficiently low temperature. However, even the presence of weak disorder or medium-ranged interactions will create an ``emulsion'' (microscopic phase coexistence) with Mott/Wigner insulating ``islands'' in between metallic ``rivers'' as proposed by Spivak and Kivelson in the context of 2DEG systems in semiconductors~\cite{SpivakKivelson2004PhRvB,Jamei:2005eb}. If so, then tuning bandwidth and/or temperature should produce a continuous variation of the metallic fraction $x$. As long as it   exceeds the percolation threshold ($x > x_c$), the system is conducting. At $x < x_c$ the metallic domains no longer connect across the system, and conduction stops, at least at $T=0$. Critical behavior now arises because we dealing with a classical percolation transition.

In this picture, the $T=0$ metal-insulator transition may occur without actually closing the insulating gap $\Delta$. Similarly, at the percolation threshold $x = x_c$ the metallic Fermi liquid at low $T$ is still stable, and consequently there is no strict divergence of the effective mass $m^*$, nor the vanishing of $T_{\mathrm{FL}}$ at the percolation threshold. We note, however, that such a (classical) percolation picture should apply only if the characteristic domain size is sufficiently larger than the characteristic correlation (or dephasing) length, therefore strictly speaking not at the lowest temperatures. However, finite-temperature variation of the transport properties should be adequately captured, as abundantly documented~\cite{dagotto2005science} in other systems featuring microscopic phase separation, such as Colossal Magneto-Resistance (CMR) manganites, for example. 

There is, however, an interesting and nontrivial feature of the percolation picture pertaining to finite temperatures. Because of its localized spins, the entropy of the Mott or Wigner--Mott insulating phase should be higher than that of the of the metal (Fermi liquid). As a result, when raising the temperature in the metallic regime, the insulating volume fraction will \emph{increase}: a manifestation of the Pomeranchuk effect~\cite{pomeranchuk1950jetp}. As a result, the resistivity should increase up to some $T = T_{\mathrm{max}}$, above which the metallic domains no longer connect, and  resistivity will decrease again~\cite{Spivak2006}, leading to resistivity maxima. This qualitative picture has been advocated in the work of Spivak and Kivelson, but no concrete prediction of the precise temperature dependence of the resistivity has been made, nor how the corresponding family of curves should scale as the transition is approached.

Interestingly, the same physical picture should in fact apply not only to Wigner--Mott transitions in semiconductors, but also the to conventional Mott transition, provided that there exists a well-defined metal-insulator phase coexistence region around the Mott point. Indeed, recent work on Mott organics~\cite{Pustogow:2021npj} revealed precisely such a phase coexistence region, albeit only at very low temperatures of the order of at most few percent of the (bare) Fermi energy. Here, careful  theoretical modeling~\cite{Pustogow:2021npj} firmly established the validity of the percolation picture, but only within a well-defined phase coexistence region. In contrast, in all the systems studied (2DEG, Mott organics, moir\'e), the pronounced resistivity maxima persist even much further onto the metallic side, where $T_{max}$ can reach a substantial fraction of $T_F$, where phase coexistence is very unlikely. Furthermore, recent experimental work on Mott organics by Kanoda and collaborators demonstrated~\cite{kanoda2019disorder,kanoda2020griffiths} the extreme fragility of such a phase coexistence region to disorder, as generally expected in 2D systems~\cite{imryma1976prl}. Nevertheless, it is extremely useful to have an independent experimental method to distinguish the phase coexistence region (where percolative effects are likely) from the regimes where a more uniform electron fluid/solid resides. The possibility to do so was spectacularly demonstrated in the context of Mott organics. In the next section we discuss how the dielectric response can tell which mechanism (quasiparticle destruction or percolation) is at play in a given regime. 

We briefly mention that percolation effects have been also discussed in the context of spinon theory in a  recent paper~\cite{Sunghoon2022}, which does require however significant disorder. On the other hand, the Spivak-Kivelson theory does not require disorder as the micro-emulsion of insulators and metals can be self-generated. This seems to be more in line with the experiments of Section~\ref{Sec:Experiments}: at least the 2DEGs~\cite{kravchenko2019} and the Mott organics~\cite{dressel2020advances} are displaying Mott criticality in the cleanest samples possible. It is therefore very plausible that most universal features observed in all critical Mott systems are not the result of disorder, but are instead the inherent manifestations of strong correlation physics.

\section{Interpreting Resistivity Maxima} 
\label{Sec:Dielectric}

As we have seen from our brief theory overview above, several scenarios were proposed, with sometimes similar predictions for characteristic features seen in experiments. A notable example is the  clear emergence of the resistivity maxima on the metallic side, at a temperature $T = T_{max} \gtrsim T_{FL}$, which is seen to decrease towards the transition. What is its physical content? The three theoretical pictures propose very different physical perspectives on what goes on here. 

As we mentioned in Section~\ref{Sec:SpinonTheory}, spinon theory~\cite{Senthil:2008ki} predicts the presence of resistivity maxima only in $d=3$, but not in $d=2$.  
However, robust resistivity maxima are clearly seen all the material systems of Section~\ref{Sec:Experiments}. 
An understanding of the resistivity maxima must therefore come from either the DMFT perspective on Mott physics, or from the percolative scenario. Both mechanisms provide reasonable albeit very different routes to explain the resistivity maxima. How should one distinguish them and thus identify the precise mechanism at play in a given system? Luckily, important clues were provided by recent experiments on Mott organics~\cite{Pustogow:2021npj}. Here one finds two distinct regimes, both featuring similar resistivity maxima, but with very different dielectric response. One such regime is corresponds to the (spatially inhomogeneous) metal-insulator phase coexistence region, where colossal enhancement of dielectric response has been found. The other regime was found further on the metallic side, where a dramatic drop and a change of sign the dielectric constant signaled thermal destruction of coherent quasiparticles due to strong correlation~effects. 

In the following we show how general scaling arguments can be used within each of the two proposed scenarios, to demonstrate the general robustness of these trends, thus providing a new window of what precisely goes on near the Mott point. 

\subsection{Resistivity Maxima from Thermally Destroying Coherent Quasiparticles}

Both experiments and theory provide evidence that a strongly correlated Fermi liquid forms on the metallic side of the Mott point, with a characteristic ``Brinkman--Rice'' (BR) energy scale 
$T_{BR} \sim 1/m^*$, which decreases towards the transition, thus characterizing the heavy quasiparticles. Inelastic electron-electron scattering increases with temperature, eventually leading to the thermal destruction of the quasiparticles around $T \sim T_{BR}$, and the associated modification of both the single particle (ARPES) spectra and the optical conductivity. At higher temperatures, transport assumes incoherent character, which can no longer be understood in terms of the quasiparticle picture or Fermi Liquid ideas alone. Its precise form generally depends on band filling and the correlation strength, but more precise predictions require a specific microscopic model and a theoretical picture. 
%

%
Concrete and quantitative results, in this regime, were given by DMFT theory, which provided first insight into the origin of the resistivity maxima in certain Mott organic materials at half-filling, as well as in certain oxides. Subsequent DMFT studies stressed that the characteristic temperature scale for the resistivity maxima indeed tracks the BR scale of the quasiparticles ($T_{max} = T_{BR}$), while preserving the functional form of the resistivity curves across this coherence-incoherence crossover. This revealed the scaling behavior of the resistivity curves in the correlated metallic regime, with a universal scaling function of $T/T_{max}$. The predicted scaling behavior has been confirmed by a number of experiments on various systems~\cite{Radonjic:2012gm, shashkin2020manifestation,moon2020quantum}, displaying even quantitative agreement with the theoretical scaling function, with no adjustable parameters.

Further optical and dielectric studies~\cite{Pustogow:2021npj} in Mott organics also confirmed the predicted destruction of the Drude peak around $T_{BR}$, again signaling the thermal destruction of quasiparticles. They established that it dramatically affects not only DC transport, but also the dielectric response, which in this metallic regime is seen to display a dramatic drop from moderate positive values at $T > T_{drop}\sim T_{max}$ to very large but negative values at $T < T_{drop}\sim T_{max}$. These studies, combining experiments and DMFT theory, have firmly established that the dielectric response can be used to directly reveal the thermal destruction of quasiparticles around the BR temperature. 

In the following, we extend the systematic studies of Ref.\cite{Radonjic:2012gm}, to stress that within DMFT both DC transport and the dielectric response display the characteristic crossover behavior across $T_{BR}$, and the associated scaling behavior upon approaching the Mott point. To do this we calculate the dielectric function $\epsilon_1$ as a function of temperature and interaction $U$, using the same setup as in our recent work~\cite{Pustogow:2021npj}. For simplicity, we focus on a simple semi-circular band model at half filling, and carry out DMFT calculations using the standard CTQMC impurity solver with the the Maximum Entropy method for analytical continuation to the real axis. Just as in Ref.\cite{Radonjic:2012gm}, once we get the single particle self energy from our DMFT equations, we calculate the real part of optical conductivity $\sigma_1 (\omega)$  from the standard Kubo formula, and the imaginary part of optical conductivity $\sigma_2(\omega)$ using the  Kramers-Kroning tranform; the (complex) dielectric function is then obtained via~\cite{Economou}
\begin{linenomath}
\begin{equation}
    \epsilon(\omega)=1+4\pi i\frac{\sigma(\omega)}{\omega}. 
\end{equation}
\end{linenomath}



\begin{figure}[H]
\includegraphics[width=0.96\textwidth]{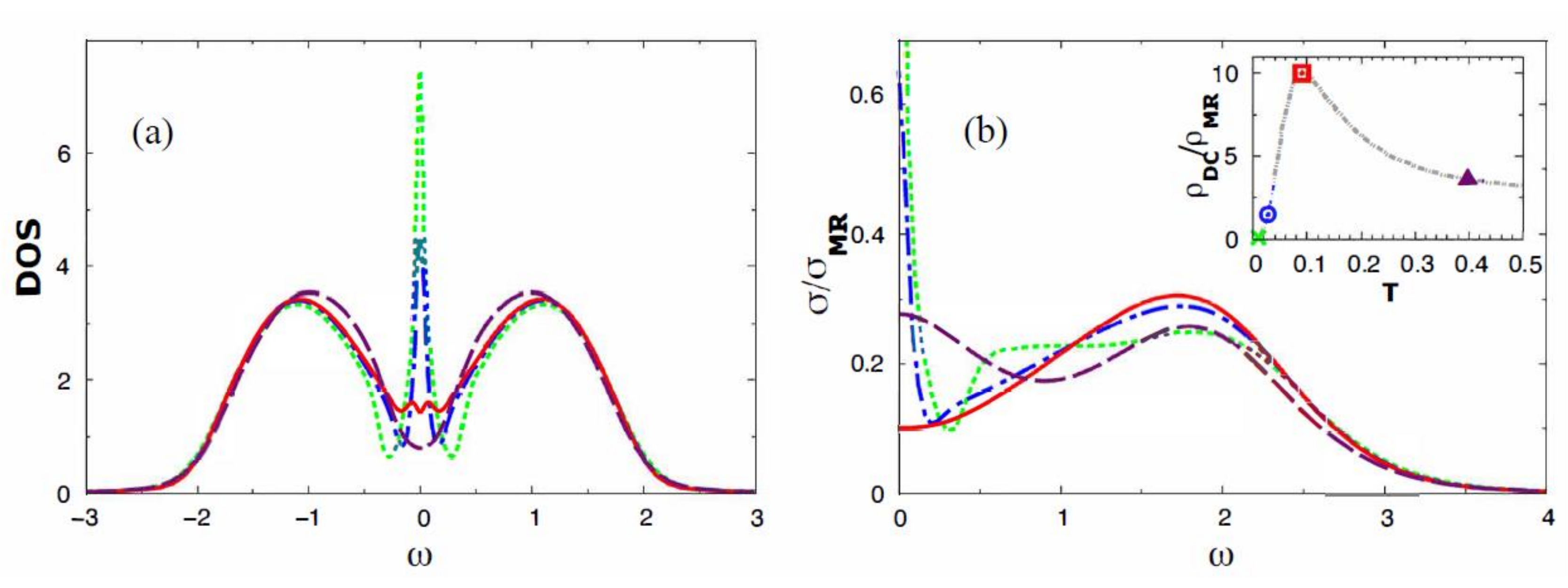}
\caption{(\textbf{a}) DMFT results for the evolution of the single-particle Density of States (DOS) for several values of the temperature ~(reprinted with permission from Ref. \cite{Radonjic2010prb} Copyright 2010 American Physical Society), as well as (\textbf{b}) that of the optical conductivity, in the strongly correlated metallic regime. Different colors correspond to the four distinctive transport regimes (inset in (\textbf{b})). DOS features a distinct quasiparticle peak at low temperatures, which is thermally destroyed at temperature $T_{max}= T_{BR} \sim (m^*)^{-1}$, where the resistivity (inset of right panel) reaches a maximum. The optical conductivity displays the corresponding suppression of the low-frequency Drude peak around the same temperature.}
\label{Fig:DMFTdos}
\end{figure}

\begin{figure}[H]
\includegraphics[width=1\textwidth]{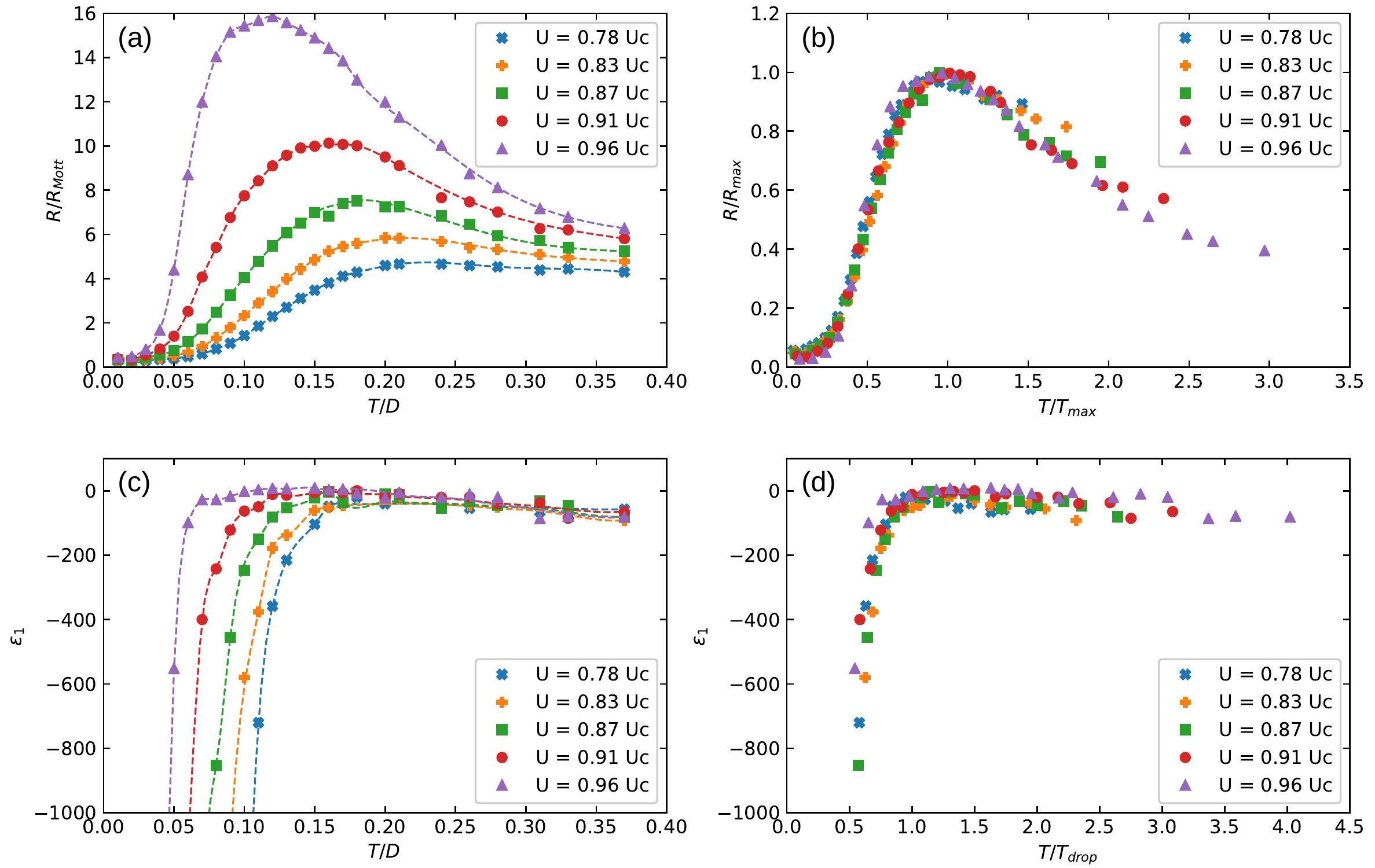}
\caption{(\textbf{a}) DC resistivity as a function of temperature for several interaction strengths. (\textbf{b}) Scaled resistivity curves. (\textbf{c}) Real part of dielectric function $\epsilon_1$ at $\omega/D=0.01$, as a function of temperature for several interaction strengths. (\textbf{d}) Scaled dielectric function curves. Results are obtained for a half-filled Hubbard model solved within DMFT.} 
\label{Fig:Dielectric1}
\end{figure}

\begin{figure}[H]
\includegraphics[width=1\textwidth]{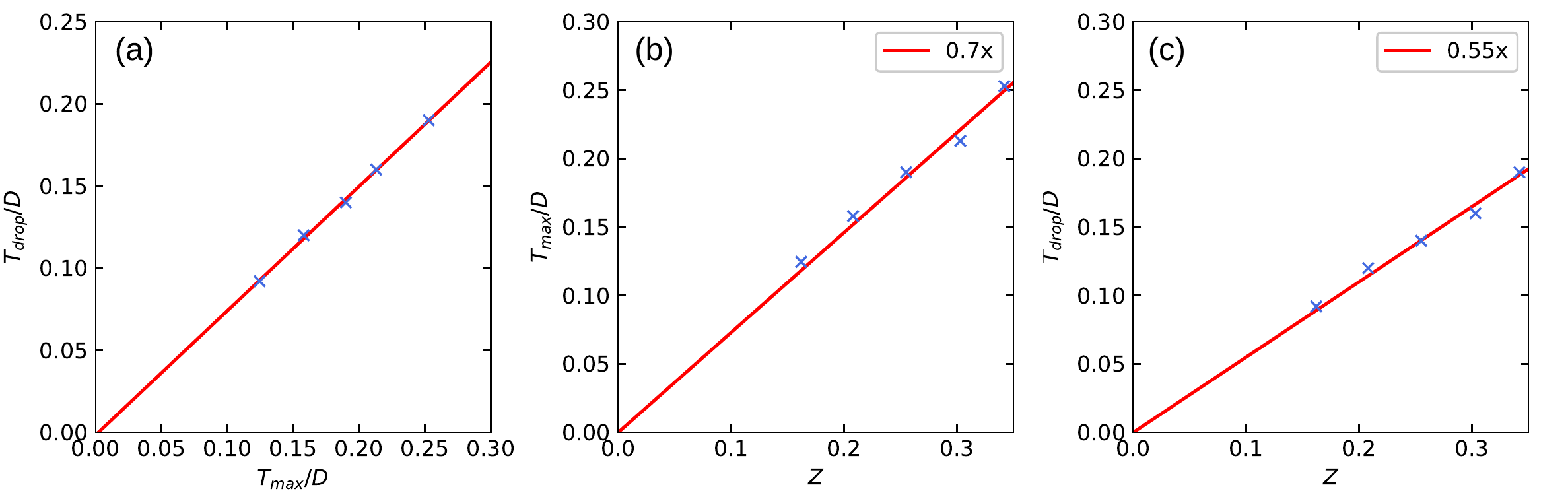}
\caption{(\textbf{a}) $T_{drop}$ as a function of $T_{max}$. (\textbf{b}) $T_{max}$ as a function of $Z$. (\textbf{c}) $T_{drop}$ as a function of $Z$. } 
\label{Fig:Dielectric2}
\end{figure}

The results for the single-particle density of states and the optical conductivity are shown in Figure~\ref{Fig:DMFTdos}, and results for the DC transport and the low-frequency dielectric response are displayed in Figure~\ref{Fig:Dielectric1}, for the parameter range corresponding to the correlated metallic phase ($ U \lesssim U_{c1} $). Here panels (a) and (b) reproduce the results of Ref.~\cite{Radonjic:2012gm}, showing the characteristic scaling behavior of the resistivity maxima near the Mott point. The analogous behavior for the dielectric function  $\epsilon_1$ is shown in panels (c) and (d), firmly establishing that the observed crossover behavior assumes a universal scaling form in the correlated metallic regime. The notion that the thermal destruction of quasiparticles lies behind both phenomena is seen even more clearly in Figure~\ref{Fig:Dielectric2}, where we show how  the scale $T_{max}$ for the resistivity maxima, and the scale $T_{drop}$ of the dielectric response, both scale with the quasiparticles weight $Z = m/m^*$, as the transition is approached. 

These results establish a way to experimentally recognize the thermal destruction of quasiparticles as a dominant mechanism behind the resistivity maxima within a correlated but uniform metallic phase. Since the correlation processes captured by DMFT are essentially {\em local} (i.e. ``ultraviolet, UV''), these effects should not display significant dependency on spatial dimensional. Indeed, experiments have shown that similar resistivity maxima are seen within correlated metallic phases both in 2D and in 3D systems. 

Note, however, that DMFT predicts very different behavior closer to the Mott point, specifically within the phase coexistence region at $T < T_c$ and $U_{c1} < U < U_{c2}$. Here, just as around any first-order phase transition line, we expect hysteresis phenomena and inhomogeneous phase separation, where metallic and insulating domains coexist on a nano-scale. As stressed in the seminal work by Spivak and Kivelson~\cite{SpivakKivelson2004PhRvB}, thermal effects can modify the relative volume fraction of the two coexisting phases, producing under appropriate conditions the characteristic resistivity maxima. In recent work motivated by experiments, a microscopic ``hybrid-DMFT'' approach was developed~\cite{Pustogow:2021npj} to quantitatively describe this regime in the context of Mott organics, resulting in spectacular agreement with experiments. In the following section, however, we wish to stress that the {\em qualitative} aspects of this regime display a number of universal scaling features, which can be precisely understood from the perspective of percolation theory.


%
\subsection{Percolation Scenario Due to Phase Coexistence}

To focus on the universal scaling aspects of percolative processes within the metal-insulator phase coexistence region, we follow the seminal ideas of Efros and Shklovskii~\cite{Efros1976}, and set up a two-component random resistor network model, with characteristic low-frequency form for the (complex) conductivity for each component: 
\begin{linenomath}
\begin{equation}
\begin{split}
    \sigma_I &= \sigma_I^o \text{exp}(-\Delta/T)-iC\omega,
    \\
    \sigma_M &= \frac{\sigma_M^o}{1-i\omega\tau}.
\end{split}
\end{equation}
\end{linenomath}

Here we assumed activated DC transport for the insulating component, with capacitance $C$ and a standard Drude form for the conducting component, with finite DC conductivity $\sigma_M^o$. To leading order near the percolation point, we ignore the $T$-dependence of  $\sigma_M^o$, $\sigma_I^0$, and $C$, since the dominant effects come from the variation of the respective volume fractions, and the activated form of insulating transport. The temperature is expressed in the units of the activation energy $\Delta$, which is also taken to be a constant. 
The corresponding expressions for the (complex) dielectric functions of the two components are given by:
\begin{linenomath}
\begin{equation}
\begin{split}
    \epsilon_I &= 1+4\pi C +\frac{4\pi i}{\omega} \sigma_I^o \text{exp}(-\Delta/T),
    \\
    \epsilon_M &= 1-4\pi\tau\sigma^o_M+\frac{4\pi i}{\omega}\sigma^o_M.
\end{split}
\end{equation}
\end{linenomath}

Here we ignored the capacitance of the metallic domains, which can be neglected if $\tau\sigma^0_M/C \gg 1$. 

\begin{figure}[H]
\includegraphics[width=0.96\textwidth]{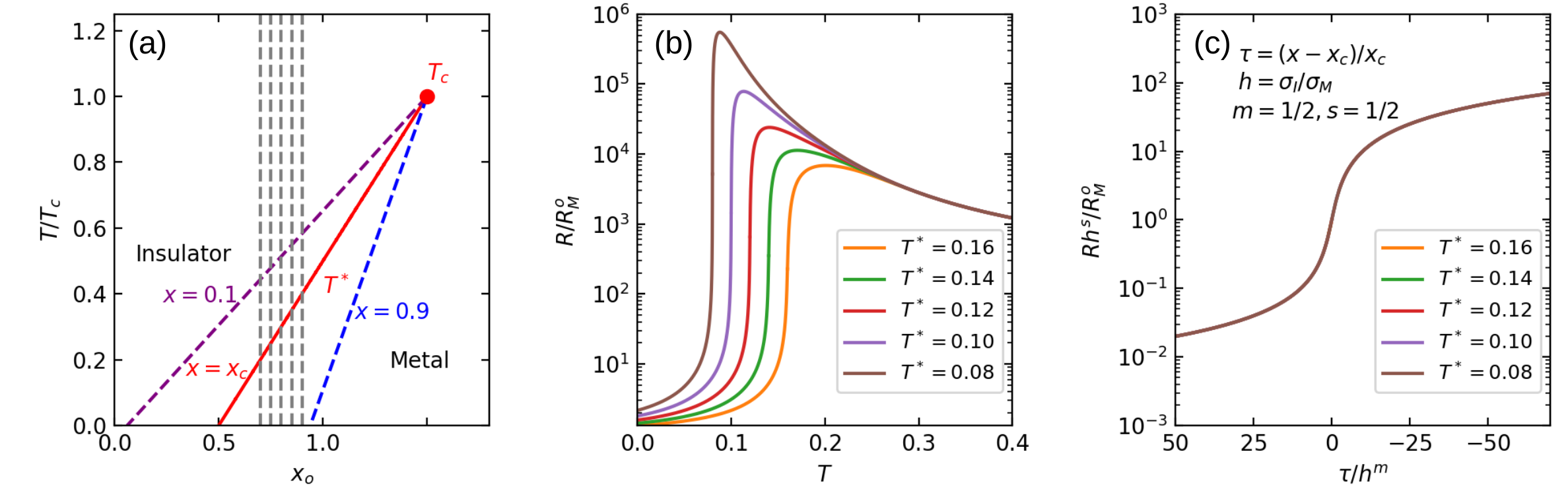}
\caption{(\textbf{a}) The red line is $x=x(T^*)$. For $T$ larger than the blue dashed line, $x=0$. We calculate the percolation results along the grey dashed lines. (\textbf{b}) $R/R^o_M$ as a function of $T$ for different $T^*$. \mbox{(\textbf{c}) Scaled} resistivity curves. } 
\label{Fig:Dielectric3}
\end{figure}

Such a random resistor network model is appropriate for any percolating two component metal-insulator system. To describe formation of the resistivity maxima, an additional physical condition has to be met, as emphasized by Spivak and Kivelson in the context of Wigner--Mott transitions, but which is in fact valid for any Mott-like  system in general. As we mentioned before, this ``Pomeranchuk effect''~\cite{pomeranchuk1950jetp} requires that the first-order line (and the entire phase coexistence region) be ``tilted'' towards the metal, so that the higher-entropy phase emerges at higher temperatures. To schematically represent such a situation we assume that, within the phase coexistence region, the volume fraction $x$ of the metallic component {\em decreases} with temperature. As an illustration, we take the following simple~model:
\begin{linenomath}
\begin{equation}
\begin{split}
x(x_o,T)=x_c + \frac{1}{2}\text{tanh}[\frac{x_o-x^*(T)}{w(T)}],
\end{split}
\end {equation}
\end{linenomath}
where $x_c$ represents the percolation threshold, $w(T)=a(T_c-T)/T_c$ defines the width of the coexistence region, and $x^*(T)=x_c + b(T/T_c)$, as illustrated in Figure~\ref{Fig:Dielectric3}a, for  $a=0.4$ and $b=1$. In this model, the parameter $x_o$ controls the metallic volume fraction at $T=0$, which decreases at $T > 0$, and reaches the percolation threshold $x = x_c$ at $T = T^* (x_o) = T_c (x_o - x_c)/b$. Physically, the the DC resistivity will first increase with $T$ as the metallic volume fraction decreases. Past percolation threshold, however, the metallic domains no longer connect.  Transport then assumes insulating (activated) form, resulting in subsequent resistivity decrease at $T > T^*$, and the emergence of resistivity maxima around $T \sim T^*$. Similarly, the dielectric constant $\epsilon_1$ grows (diverges) as the percolation threshold is approached from the insulating side, due to the formation of large metallic clusters with increased polarizability. On the metallic side, however, it displays a rapid decrease, dropping to large negative values within the metallic phase. As a result, dielectric response displays colossal enhancement around the percolation threshold, a phenomenon that can be viewed as a smoking gun for percolative charge dynamics. 

\begin{figure}[H]
\includegraphics[width=1\textwidth]{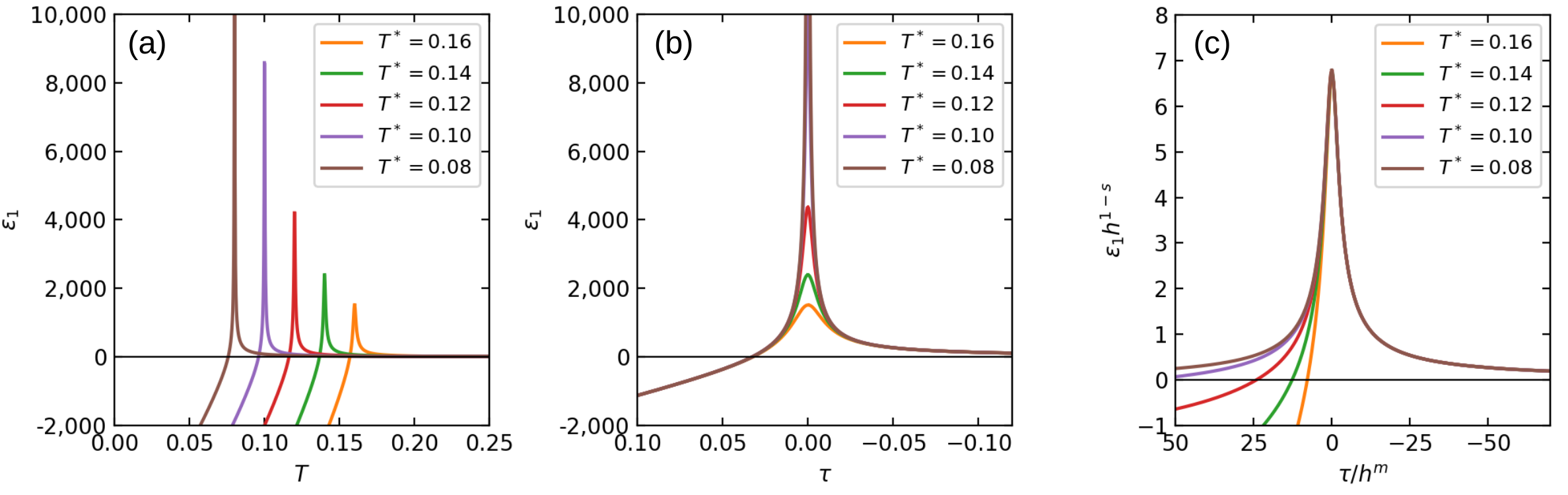}
\caption{(\textbf{a}) The dielectric constant $\epsilon_1$ as a function of $T$ for different $T^*$. (\textbf{b}) $\epsilon_1$ as a function of $\tau$ for different $T^*$. (\textbf{c}) Scaled dielectric function curves.  }
\label{Fig:Dielectric4}
\end{figure}

\textls[-25]{To illustrate these ideas, we use the Effective Medium Approximation (EMA) for percolation, which solves the following nonlinear equations for the complex dielectric function:}
\begin{linenomath}
\begin{equation}
\begin{split}
    x(\frac{\epsilon_M-\epsilon}{\epsilon_M+(z/2-1)\epsilon})+(1-x)(\frac{\epsilon_I-\epsilon}{\epsilon_I+(z/2-1)\epsilon})=0,
\end{split}
\end{equation}
\end{linenomath}
 and for illustration selected $\sigma_M^o / \sigma_I^o = 100, \tau\sigma^o_M=1000$, $C=1$, $T_c/\Delta =0.4$, and $z=4$ corresponding to 2D transport ($x_c = 0.5$). Precisely the anticipated behavior is observed from numerically solving the EMA equation for the corresponding DC resistivity $R = \sigma^{-1}$, as shown in Figure~\ref{Fig:Dielectric3}b. Here, we select several values of $x_o$, corresponding to $x > x_c$ (low temperature metallic regime), and plot the resistivity as a function of temperature (following dashed lines in Figure~\ref{Fig:Dielectric3}a. We observe distinct resistivity maxima around the temperature $T^* (x_o)$ corresponding to the percolation threshold. Note how the maxima become sharper and sharper as $T^*$ is reduced, corresponding to the exponential (activated) decrease of the ``field'' $h \sim \exp\{ -\Delta/T^*\}$. The expected behavior is also seen in dielectric response, as shown in  Figure~\ref{Fig:Dielectric4}a, where we observe sharp maxima at $T \sim T^*$. Here again we see the increased ``rounding'' of these maxima at higher $T^*$, corresponding to larger $h(T^*)$. This behavior can be seen even more clearly in Figure~\ref{Fig:Dielectric4}b, where $\epsilon_1$ is plotted as a function of the reduced concentration $\tau = (x(T)-x_c)/x_c$, which vanishes at $T = T^*$. 
 
 These qualitative trend can be even more rigorously described within the scaling theory for percolation, where the DC conductivity as well as the $\omega = 0$ dielectric constant are known to satisfy the following scaling relations:
\begin{linenomath}
\begin{equation}
\sigma_1(\tau, h) = \sigma_M^o h^s F_{\sigma} (\tau / h^m); \; \; \; \epsilon_1(\tau, h) = h^{s-1} F_{\epsilon}(\tau / h^m),
\end{equation}
\end{linenomath}
where $\tau = (x(T)-x_c)/x_c$ measures the distance to the percolation threshold, and \linebreak $h=\sigma_I / \sigma_M$ plays a role of the ``symmetry breaking field'', which leads to the rounding of the transition. The critical exponents $s$ and $m$, as well as the crossover scaling functions \mbox{$F_{\sigma}$ and $F_{\epsilon}$} are universal quantities within percolation theory. To illustrate this scaling behavior  within EMA, we collapse the family of resistivity curves by plotting $R h^s$ as a function of $\tau/h^m$, as shown in Figure~\ref{Fig:Dielectric3}c, and $\epsilon_1 h^{1-s}$ as a function of s a function of~$\tau/h^m$, as shown in Figure~\ref{Fig:Dielectric4}c. Note how a perfect scaling collapse is observed here, but only around the peak of the dielectric response, i.e. only close to the percolation threshold. Such behavior is, in fact, not surprising, since we expect scaling phenomena to arise only within a given critical region, and not further away from the critical point. 

We should stress again that all our qualitative results are rigorously valid within general percolation theory for our two-component phase coexistence model, and EMA was simply used as an illustration. EMA correctly captures the general crossover phenomena associated with percolation, but only introduces approximate values for the critical exponents $s_{EMA}=0.5$ and $m_{EMA} = 0.5$, which are otherwise know even more accurately from numerical simulations. These details, however, are not of direct relevance for our purposes. What is important is the result that, within our ``Pomeranchuk'' model for phase coexistence, the percolation scenario predicts distinct resistivity maxima but also striking colossal dielectric anomalies, at the same temperature scale of $T = T^*$ which decreases towards the MIT. This behavior is in distinct contrast to the behavior we found from the DMFT picture of a correlated but uniform metallic phase, which also leads to resistivity maxima, but very different behavior of the dielectric response. This observation, which was quantitatively validated in recent experiments on Mott organics~\cite{Pustogow:2021npj}, thus reveals a distinct criterion to settle the long-lasting controversies between the origin of the resistivity maxima in different systems. 

\section{Conclusions} 
\label{Sec:Conclusion}

In this paper we discussed three different classes of physical systems which all display very similar phenomenology expected for Mott-like metal-insulator transitions. We stressed that most qualitative features are clearly seen in all these examples, including the continuous decrease of the characteristic energy scales $T_{FL}$, $T_{BR} = T_{max}$, $T_o$, $\Delta$ towards the transition, the phenomenon of quantum critical scaling seen in transport, as well as the emergence of distinct resistivity maxima on the metallic side. These observations, which are starting to portray a robust and consistent phenomenology of Mott criticality, is putting serious constraints on theory. We discussed which of these features seem compatible with various proposed theoretical pictures of the Mott point, and which ones do not. 

In the final section of this paper we also presented new theoretical results, which open the possibility to precisely determine, from experiments, which mechanism dominates in which regime. We argue that the {\em dielectric response} offers unique insights, which so far have not been appreciated enough, as a powerful tool to distinguish between different phase coexistence and the thermal destruction of quasiparticles.

A class of issues we did not discuss in any detail in this paper is the (explicit) role of disorder around the Mott point. Given the fact that new classes of ultra-clean material are starting to emerge, with even more pronounced salient features of Mott criticality, it is becoming possible to plausibly minimize the role of disorder on experimental grounds. On the other hand, new experimental efforts are starting~\cite{kanoda2019disorder,kanoda2020griffiths} to emerge in the opposite direction: to systematically add and to control the level of disorder, for example by high-energy X-ray irradiation. These fascinating research directions are guaranteed to open entirely new chapters in the study of metal-insulator quantum criticality. This will require theorists to rekindle the efforts to understand the interplay of strong correlation with disorder~\cite{miranda2005disorder}, and perhaps to develop new ideas in the process.

\vspace{6pt} 



\authorcontributions{Y.T. carried out theoretical calculations and performed the analyses. V.D. and L.R. designed the project.  All authors discussed the data, interpreted the results, and wrote the paper. All authors have read and agreed to the published version of the manuscript. }

\funding{This work was supported by the NSF Grant No. 1822258, and the National High Magnetic Field Laboratory through the NSF Cooperative Agreement No. 1644779 and the State of Florida. LR acknowledges support by the Swiss National Science Foundation via Ambizione grant PZ00P2\_174208.}

\institutionalreview{Not applicable.
}
\informedconsent{Not applicable.

}
\dataavailability{
Data are provided: \url{https://github.com/yutingtanphysics/} (accessed on 3 May 2022) How-to-Recognize-the-Universal-Aspects-of-Mott-Criticality.} 
\acknowledgments{We thank Pak Ki Henry Tsang for providing technical support in doing computer cluster calculations.} 

\conflictsofinterest{The authors declare no conflict of interest. The funders had no role in the design of the study; in the collection, analyses, or interpretation of data; in the writing of the manuscript, or in the decision to publish the~results.}

\begin{adjustwidth}{-\extralength}{0cm}

\reftitle{References}

\end{adjustwidth}
\end{document}